\documentclass[pra,notitlepage,groupaddress,nofootinbib,superscriptaddress,longbibliography,twocolumn]{revtex4-1}

\AtBeginDocument{%
    \newwrite\bibnotes
    \def\bibnotesext{Notes.bib}
    \immediate\openout\bibnotes=\jobname\bibnotesext
    \immediate\write\bibnotes{@CONTROL{REVTEX41Control}}
    \immediate\write\bibnotes{@CONTROL{%
    apsrev41Control,author="08",editor="1",pages="1",title="0",year="1"}}
     \if@filesw
     \immediate\write\@auxout{\string\citation{apsrev41Control}}%
    \fi
}%

\usepackage{amsmath,amsfonts,amssymb,graphicx,color,times,bbm,psfrag,dsfont}
\usepackage{bbold}
\usepackage{mathrsfs}
\usepackage{slashed}

\usepackage{float}
\usepackage[caption=false]{subfig}
\usepackage{tikz}
\usepackage{pgfplots}
\usepackage[theorems,skins]{tcolorbox}
\newtcolorbox{mymathbox}[1][]{colback=white, ams gather, outer arc=0pt, #1}
\usepackage{mathrsfs}
\usepackage{epstopdf}
\usepackage{multirow}
\usepackage{gensymb} 

\usepackage{xr}
\usepackage{zref-xr}
\usepackage{chngcntr}
\usepackage{siunitx}
\usepackage{mathtools} 

\definecolor{myblue}{rgb}{0.1,0.24,0.6}
\definecolor{myred}{rgb}{0.6,0.1,0.2}
\usepackage[colorlinks=true, citecolor=myblue, linkcolor= myblue, urlcolor=myblue,bookmarks=true]{hyperref}

\newcommand{\Ai}[1]{ {\mathrm{Ai} } \left ( #1 \right ) } 
\newcommand{\Bi}[1]{ {\mathrm{Bi} } \left ( #1 \right ) }

\newcommand{\ket}[1]{\left \rvert #1 \right \rangle}
\newcommand{\bra}[1]{\left \langle #1 \right \rvert}
\newcommand{\overlap}[2]{\left \langle #1 \rvert  #2  \right \rangle}
\newcommand{\expect}[1]{\left \langle #1 \right \rangle}

\def \beq{\begin{equation}}
\def \eeq{\end{equation}}
\def \bse{\begin{subequations}}
\def \ese{\end{subequations}}
\def \bea{\begin{eqnarray}}
\def \eea{\end{eqnarray}}
\def \bem{\begin{pmatrix}}
\def \eem{\end{pmatrix}}
\def \bs{\boldsymbol}

\begin{document}

\title{Transport across twist angle domains in moir\'e graphene}
\author{Bikash Padhi }
\email{bpadhi2@illinois.edu}
\affiliation{Department of Physics and Institute for Condensed Matter Theory, University of Illinois at Urbana-Champaign, Urbana, IL 61801, USA}
\author{Apoorv Tiwari}
\affiliation{Department of Physics, University of Z\"urich, Winterthurerstrasse 190, 8057 Z\"urich, Switzerland}
\affiliation{Condensed Matter Theory Group, Paul Scherrer Institute, CH-5232 Villigen PSI, Switzerland}
\author{Titus Neupert}
\affiliation{Department of Physics, University of Z\"urich, Winterthurerstrasse 190, 8057 Z\"urich, Switzerland}
\author{Shinsei Ryu}
\affiliation{Kadanoff Center for Theoretical Physics and James Franck Institute, University of Chicago, Chicago, IL 60637, USA}

\begin{abstract}
Many of the experiments in twisted bilayer graphene (TBG) differ from each other in terms of the details of their phase diagrams. Few controllable aspects aside, this discrepancy is largely believed to be arising from the presence of a varying degree of twist angle inhomogeneity across different samples. Real space maps indeed reveal TBG devices splitting into several large domains of different twist angles. Motivated by these observations, we study the quantum mechanical tunneling across a domain wall (DW) that separates two such regions.
We show that the tunneling of  the moir\'e particles can be understood by the formation of an effective step potential at the DW. The height of this step potential is simply a measure of the difference in twist angles.
These computations lead us to identify the global transport signatures for detecting and quantifying the local twist angle variations. In particular, 
Using Landauer-B\"uttiker formalism we compute single-channel conductance ($dI/dV$) and Fano factor for shot noise (ratio of noise power and mean current). A zero-bias, sub-meV transport gap is observed in the conductance which scales with the height of the step potential. One of the key findings of our work is that transport in presence of twist angle inhomogeneity is ``noisy", though sub-Poissonian. In particular, the differential Fano factor peaks near the van Hove energies corresponding to the domains in the sample. The location and the strength of the peak is simply a measure of the degree of twist angle inhomogeneity.
\end{abstract}

\maketitle

\tableofcontents
\onecolumngrid
\vspace{1cm}
\twocolumngrid

\clearpage

\onecolumngrid
\begin{figure*}
\centering
\includegraphics[width=0.40 \textwidth]{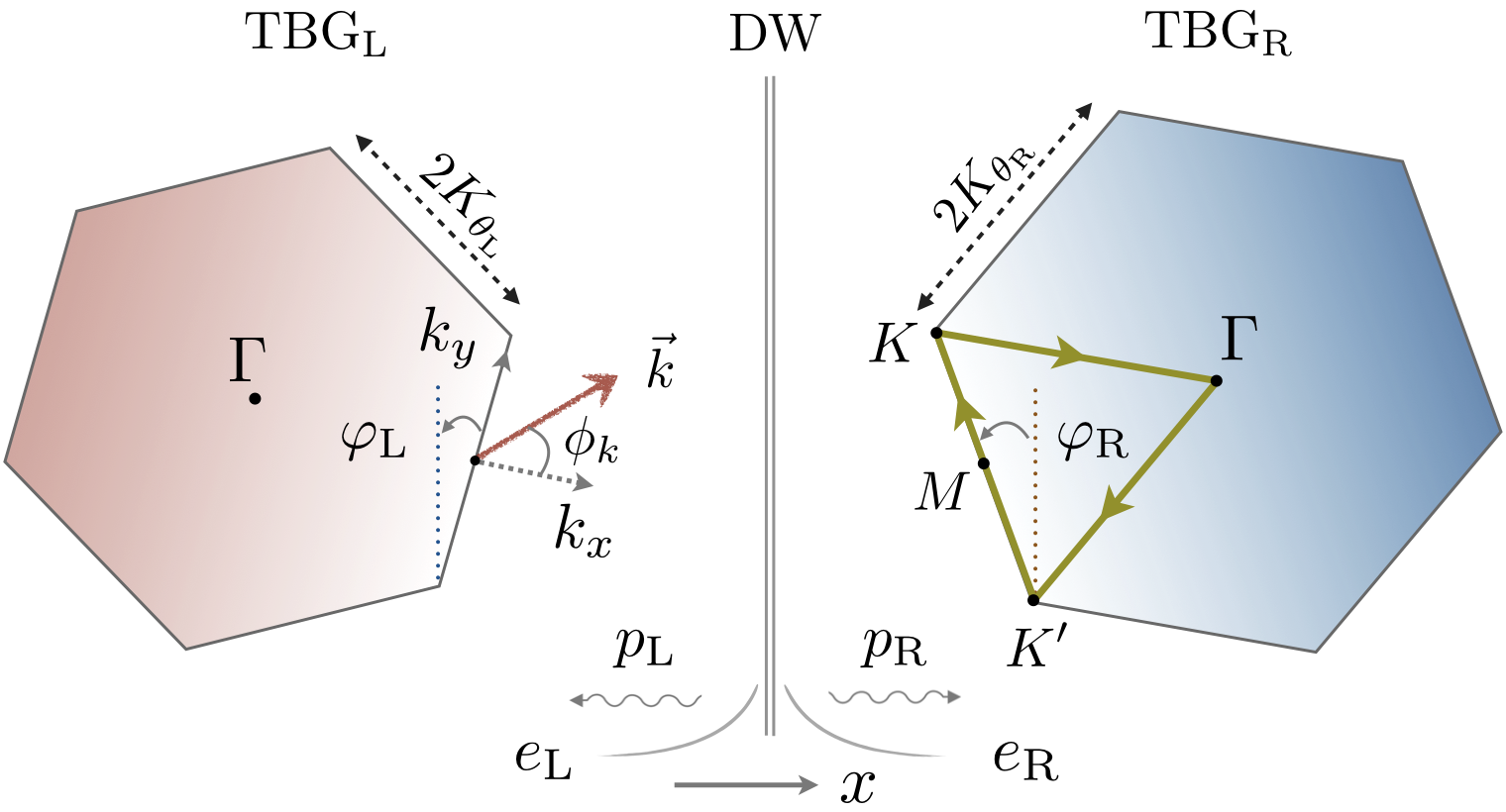}  \,\, 
\includegraphics[width=0.30 \textwidth]{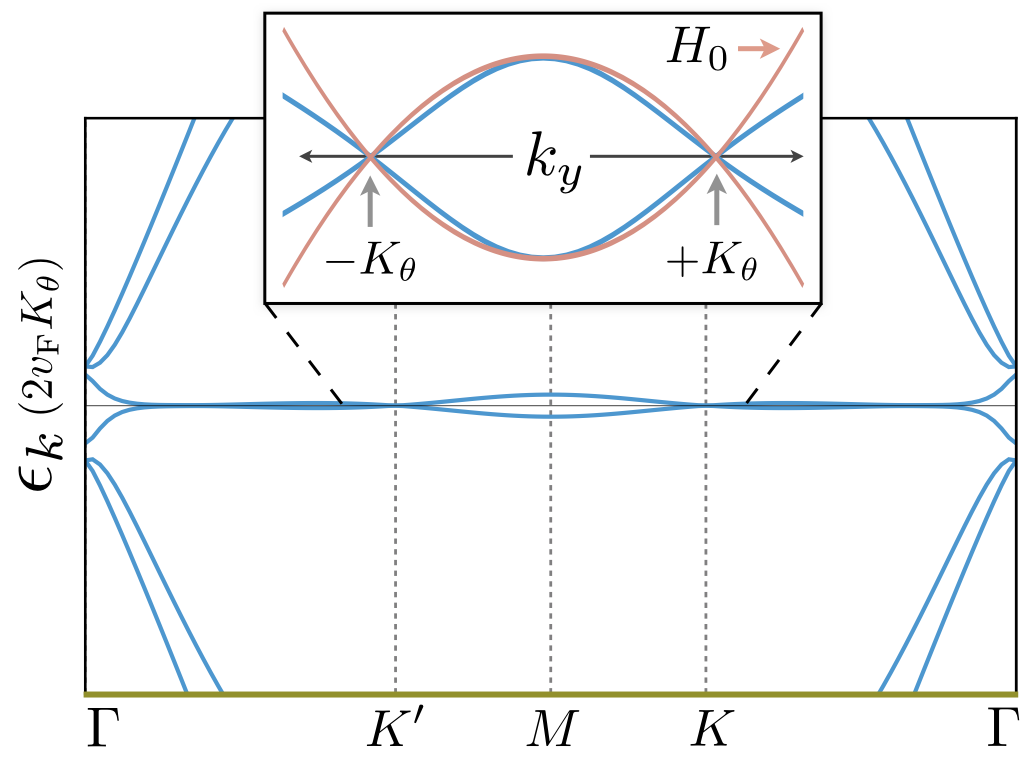} \,\,\, 
\includegraphics[width=0.26 \textwidth]{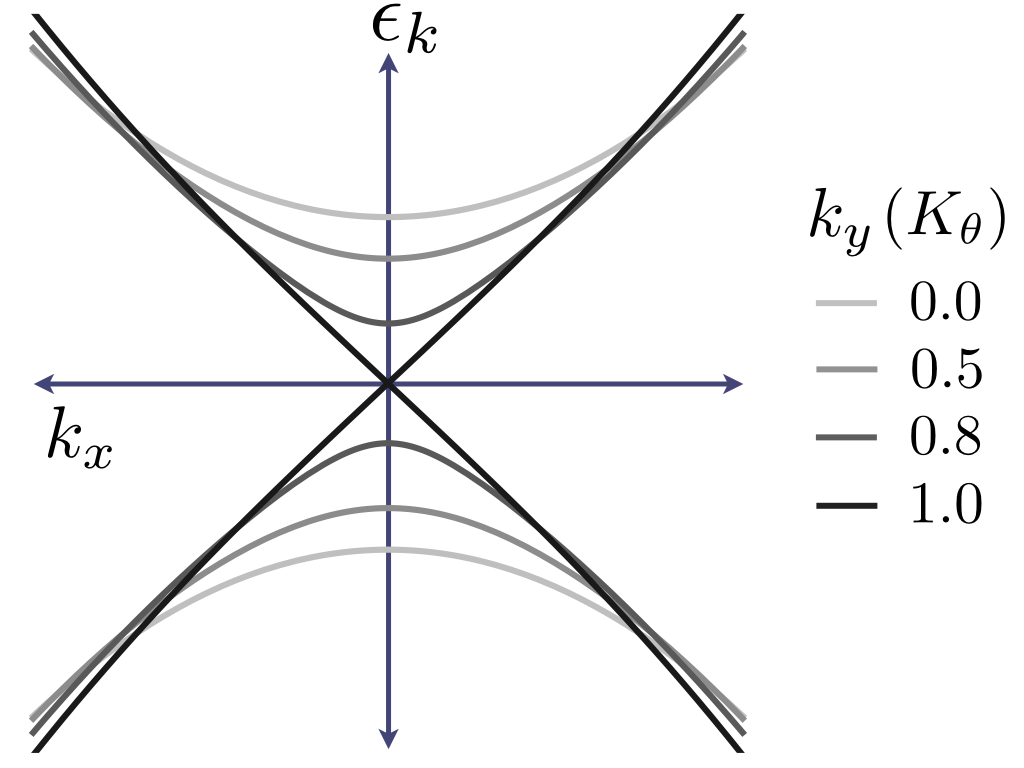}
\caption{
Schematic of the system: (Left)  A TBG sample with two different twist angles ($\theta_{\mathrm{L} }, \theta_{\mathrm{R} }$) across a domain wall along the $y$-axis. We analyze tunneling characteristics of incident moir\'e electrons from the left, with momentum $\vec k$ measured from the $M$ point and incident angle $\phi_k$. In TBG there always exist two evanescent modes, $e_{ \mathrm{L,R} }$, near a DW. Moir\'e electrons can tunnel when there also exist propagating modes, $p_{ \mathrm{L,R} }$. In general, the MBZs of sizes $2K_{\theta_{ \mathrm{L,R} }}$ can tilt by $\varphi_{ \mathrm{L,R} }$ angles with respect to the DW. (Middle) Using the Bistritzer--MacDonald model, the low energy moir\'e dispersion ($\theta=1.18^\circ$) is obtained along the cut (green lines) shown in the left panel. The inset compares this with the effective dispersion obtained in Eq.~\eqref{eq:EOM}. (Right) For a BZ-cut transverse to the zone boundary ($K$-$K'$ line), the dispersion (black curve) is gapless at the Dirac points ($k_y = \pm K_\theta$) but is maximally gapped (gray curve) at the $M$ point ($k_y=0$).
}
\label{fig:Schema}
\end{figure*}
\twocolumngrid

\section{Introduction}

Temperature, pressure, doping, etc., are among the most innate tuning parameters which can fundamentally transform, thereby help us understand, many condensed matter systems. 
As a result of the discovery of correlated insulation~\cite{Cao18Mott} and superconductivity~\cite{Cao18SC} in twisted bilayer graphene (TBG), twist angle has been envisaged as a conspicuously novel control parameter that can allow various layered van der Waals materials to host myriads of intriguing phases~\cite{TrilayerSpinPol, 3layer/hBN, WS2/MoSe2,MagicContWSe2, WSe2/WS2}.

Various scenarios were soon proposed in order to understand the nature of these, seemingly strongly correlated, phases. However, in doing so, addressing the role of electronic interaction has been a major challenge for theorists and experimentalists alike. Attempts to control the insulating or superconducting states by controlling interaction~\cite{liu2020tuning, Efetov2Interplay, saito2019decoupling}, or otherwise~\cite{YoungDean,EfetovSC,PasupathySTM,Caltech19,Grisha19Tlinear,EvaAndreiTBG,CascadePrinceton,CascadeMIT,Efetov2Interplay}, has led to a surprisingly large number of starkly contrasting phase diagrams of TBG. In fact, even the number of insulating regions and that of the superconducting domes in these samples (under almost equivalent external circumstances) have been widely different. Such discrepancies in the phase diagrams from one sample to another has created a major bottleneck in understanding these newly discovered states. 

Recent developments in real space imaging of TBG samples~\cite{KimReconstruction,PasupathySTM,uri2019mapping} have made it clear that, what makes each moir\'e device unique is the presence of a large amount of spatial inhomogeneity in twist angles, which is both undesirable and uncontrollable. In fact, other than direct images, indirect evidences of twist angle inhomogeneity has always lurked even in the first set of TBG samples. For instance, the presence of Fraunhofer oscillation in critical current varying with normal field~\cite{Cao18Mott,Cao18SC,YoungDean,EfetovSC} signals formation of superconducting domains coexisting with the normal state. Several devices also show variable transport characters depending on the lead location~\cite{YoungDean}. All these naturally hint at an inherent inhomogeneity in the TBG samples. Hence, a careful analysis of this new type of disorder -- henceforth to be dubbed as `twist disorder' --  is inevitable for a complete understanding of the phase diagram of TBG.

In this work, we study the transport properties of `moir\'e electrons' (low-energy quasiparticles in moir\'e materials) in TBG with twist disorder. In particular, we analyze the quantum mechanical tunneling of these quasiparticles across two (or more) domains, each with a different twist angle. We refer to this as `moir\'e tunneling'. Although a realistic TBG device possesses multiple twist angle domains (TADs) of various shapes and sizes~\cite{uri2019mapping}, for analytic tractability, we confine our study to tunneling across a single, or a few, one-dimensional domain walls (DWs) separating two semi-infinite TADs.

We analyze moir\'e tunneling for various scenarios differing in four free parameters, two of which characterize the TADs and the other two characterize the moir\'e electrons. A TAD is described by two fixed angles -- twist angle ($\theta$), and tilt angle ($\varphi$). The relative orientation of the DW with respect to the zone edge ($K$-$K'$ line) of the moir\'e Brillouin zone (MBZ) is referred here as `tilt angle', see Fig.~\ref{fig:Schema}. In real space this corresponds to the angle at which the edges of two domains meet at the DW. The remaining two parameters, that describe a moir\'e electron, are its momentum and energy. In presence of a DW, translation symmetry is broken along the direction transverse to the DW, though not along the longitudinal direction. We denote this conserved component of momentum as $k_y$ since the DW is aligned along the $y$-axis. Note that one can also choose the incidence angle ($\phi$) as an equivalent control variable. Lastly, we tune the energy ($\epsilon$) of the incident electron and obtain tunneling as a function of $\epsilon$. This simple single particle analysis of moir\'e electrons help us uncover many intriguing aspects of TBG.

\begin{center}
\textbf{Summary}
\vspace{1mm}
\end{center}

Irrespective of the value of twist angle (as long as it is of about $1^\circ$), two of the most robust features of the moir\'e bandstructure of TBG~\cite{LopesPRL, LopesPRB, MCDMoire} are the band touching at the $K$ point, or the charge neutral point (CNP), and the presence of a van Hove singularity (vHS) at the $M$ point. Our tunneling study is in fact mostly focused near these two high symmetry points. The key features of moir\'e tunneling near these points are the following.

\textit{$K$-point}: In the absence of bias or doping the physics is largely dominated by the low-energy electrons near the $K$ (or $K'$) point. In the presence of a potential barrier, the dispersion being linear near the $K$ point, many of the moir\'e tunneling characters resemble those observed in pristine graphene~\cite{ChiralTunn}, e.g. Klein tunneling~\cite{NovoselovKlein, NobelSymp, ResTunTrain}. However, we show that Klein tunneling cannot occur for tunneling across a twist angle DW. In fact, one can easily establish that Dirac particles can never tunnel to the other side of the DW. This is simply because the bandstructure ensures that a Dirac particle on one side of the DW always encounters a gap on the other side. Therefore, if the energy is not sufficiently high, the tunneling is completely prohibited, resulting in a vanishing tunneling probability. Two important consequences of this fact are -- (a) due to the absence of low-energy tunneling states,  conductance contribution from moir\'e tunneling does not posses a minimum, (b) since tunneling cannot resume until the gap is overcome, this drives a zero-bias gap in conductance.

We provide an additional discussion on the recipe to resurrect Klein tunneling in TBG-like systems. We show that if one forms a DW joining two materials featuring linear dispersions with differing slope of the cones (or Dirac speeds), one can achieve not only Klein tunneling but also an electronic equivalent of the Snell's law of refraction. This follows simply from the conservation of energy and momentum. We also show these results to remain impervious to any amount of titling of the TADs. 

\textit{$M$-point}: 
Close to half-filling of the moir\'e unit cell, one can access the electrons near the $M$ point. When the energy is close to the vHS, due to the enhanced density of states (DOS), instabilities can surface even for weak interactions. This can give rise to new phases of matter. Indeed the most interesting correlated phases in TBG are seen around the half-filling point. One of the central results of our work is that we show tunneling of the high-energy moir\'e electrons near the $M$ point is analogous to the tunneling of non-relativistic electrons across a step potential. Here, the height of the effective step potential turns out to be proportional to the difference in twist angles across the DW. Using this analogy we also establish that no matter the number of the DWs (arranged in parallel) normal tunneling of moir\'e electrons is dictated \textit{only} by the twist angle of the first and the last domain. In fact, if they happen to be the same, the tunneling probability becomes identity.

We present all these results in the following manner -- in Sec.~\ref{sec:2band} we introduce the effective model that is used for all our computations. In Sec.~\ref{sec:Tunneling} we describe the method for computing moir\'e tunneling. We then divide our analysis into two parts -- in Sec.~\ref{sec:LongiDW} we study tunneling across a DW that is parallel to the $K$-$K'$ zone boundary, called longitudinal DW, and in Sec.~\ref{sec:Tilted} we extend this study to include the effects of any finite tilt angle. Various tunneling scenarios, such as the presence of multiple DWs, or when the DW has a finite width, are detailed in the subsections therein.  In Sec.~\ref{sec:Transport} we compute the mesoscopic conductance of moir\'e electrons and discuss its key features in presence of TADs. Here we also compute the Fano response and propose that this could be used as a simpler experimental tool, as compared to more involved local measurements, to diagnose twist disorder in TBG. We summarize all our findings and conclude our discussions in Sec.~\ref{sec:Conclude}.

\section{Effective Two--Band Model}
\label{sec:2band}

For twist angles $\mathcal O(1^\circ)$, TBG can be described using a host of continuum models~\cite{LopesPRL, LopesPRB, MCDMoire, NamKoshino17, AshvinPo, FaithfulAshvin, Tarnopolsky}. A key commonality in all these models, topological aspects aside, is that they describe the two graphene layers via their low-energy Dirac descriptions at a particular chosen valley and then turn on an inter-layer coupling through a moir\'e potential. Upon increasing the moir\'e potential the lowest energy branches start developing saddle points near the $M$ point, accompanied by a gap opening which isolates these bands from the high-energy branches. For small twist angles~\cite{valid2band, ChiralTunn}, Ref.~\cite{2band} obtains a minimal model from the low-energy continuum model that captures these essential features of the lowest two bands (near the $K$ valley of the original Brillouin zone) 
\begin{align}
H_0  = m_0 \begin{bmatrix} 0 & ( \hat k^\dag )^2 -  \left( \Delta K^\dag \right)^2 \\ (\hat k)^2 -   \Delta K^2 & 0 \end{bmatrix} \,\, , \,\, m_0 =\frac{2v_\mathrm{F}^2}{15 \tilde t_\perp} .
\label{eq:2band}
\end{align}%
Here, $v_\mathrm{F} \approx 10^6$ m/s is the velocity of the Dirac electrons in pristine graphene. The mass scale $m_0$ has a mild twist angle dependence via the inter-layer coupling,  $t_\perp \approx 0.27$ eV. However, for small twist angles it can be approximated~\cite{LopesPRB} to a constant, $\tilde t_\perp \simeq 0.4 t_\perp$. Henceforth, we fix $m_0 = 1$. We define $\hat k = \hat k_x + i \hat k_y$, with $\hat k_i = - i \partial_i$ and the complex momenta in the moir\'e Brillouin zone (MBZ) are defined as $k = k_x + i k_y = |k| e^{i \phi_k}$, with the origin at the $M$ point of the MBZ. Interchanging $\hat k$ with $\hat k^\dag$ (the Hermitian conjugate of $\hat k$) results in a theory near the $K'$ valley. The above matrix is written in the basis corresponding to the sublattices $A$ and $B$ of layer 1 and 2, respectively. The Dirac points in the MBZ are located at $k = \pm \Delta K$, which are obtained by twisting the Brillouin zones of the top graphene layer by an angle $\theta$, thus $\Delta K=  K_\theta e^{i \left(\varphi + \pi/2 \right)}$, where $2K_\theta = 2K\sin(\theta/2)$ is the size of the MBZ.  $K = {4\pi}/{3a_0}$ and $a_0 \approx 0.25$ nm are the momentum space and real space lattice constants of pristine graphene, respectively. The phase arising due to a finite tilt angle $\varphi$ does not have any observable consequence in absence of a DW. Due to the hexagonal symmetry of the MBZ, we restrict the value of the tilt angle to $| \varphi| \leq \pi/6$. 

The energy dispersion obtained from the effective Hamiltonian in Eq.~\eqref{eq:2band} takes the form
\begin{align}
\epsilon^2(k) =  \epsilon_\mathrm{v}  K_\theta^2 + 2 \epsilon_\mathrm{v} |k|^2  \cos \left( 2 \phi - 2 \varphi \right) + \epsilon_0 |k|^4,
\label{eq:Dispersion}
\end{align}%
where $ \epsilon_\mathrm{v} = m_0 K_\theta^2 =  \epsilon(0)$ is the saddle point energy corresponding to a logarithmic van Hove singularity (vHS) at the $M$ point~\cite{LopesPRL}, see Eq.~\eqref{eq:vHSlog}. It is important to stress here that this model, though rudimentary, correctly captures the presence of the vHS (with respect to the Dirac point). In monolayer graphene (MLG) or in Bernal stacked bilayer graphene (BLG) the vHS lies far away from the Dirac point, thus, rendering them difficult to gate. However, the proximity of the vHS to the CNP in TBG allows one to move the Fermi surface close to the vHS with ease~\cite{EvaVHS}. This enhances the density of states, thereby amplifying the interaction, leading to various instabilities and a host of different phases.

\section{Tunneling Computation}
\label{sec:Tunneling}

We now place a one-dimensional (1D) DW at $x=0$ in the above theory and proceed to compute the tunneling across it. In this section we present the method to obtain tunneling across two TADs that are tilted at an arbitrary angle $\varphi$ with respect to the DW.

Due to the presence of a DW, translation symmetry is now broken along the $x$-direction. To the left (right) of this DW there is a TBG with a twist angle $\theta_{\mathrm{L} }$ ($\theta_{\mathrm{R} }$) and a tilt angle $\varphi_{\mathrm{L} }$ ($\varphi_{\mathrm{R} }$), see Fig.~\ref{fig:Schema}. We will assume $|\theta_{\mathrm{L} } - \theta_{\mathrm{R} } | < 1^\circ$ so that we can work with a simple one-dimensional DW. Formally, the system can be described using Heaviside $\Theta$-function as 
\begin{align}
H_{ \mathrm{DW} } = H_0(\varphi_{\mathrm{L} }, \theta_{\mathrm{L} }) \, \Theta(- x) + H_0(\varphi_{\mathrm{R} }, \theta_{\mathrm{R} }) \, \Theta( x) \,. 
\end{align}%
Note that $H_{DW}$ still has translation symmetry along the $y$-direction. Therefore, we can reduce the problem to a 1D eigen-value problem after replacing $k_x$ with $- i \partial_x$ in Eq.~\eqref{eq:2band}, 
\begin{align}
\epsilon^2 \Psi  = \left( \partial_x^4 - a \partial_x^2 -  i b \partial_x  + c  \right) \Psi  .
\label{eq:EOM}
\end{align}
The parameters appearing in the above eigen-equations are position ($x$-axis) dependent step functions since they depend on the twist and the tilt angles,
\begin{align}
\begin{split}
a(\theta, \varphi) =&\; 2 \left(K_\theta^2 \cos 2 \varphi + k_y^2 \right),  \\
b(\theta, \varphi) =&\; 4 k_y K_\theta^2 \sin 2 \varphi,   \\
c(\theta, \varphi) =&\;   k_y^4 + K_\theta^4 - 2 k_y^2 K_\theta^2 \cos 2\varphi.
\label{eq:defabc}
\end{split}
\end{align}
The values abruptly switch from $a_{\mathrm{L} }\equiv a(\theta_{\mathrm{L}},\varphi_{\mathrm{L}})$ to $a_{\mathrm{R} }\equiv a(\theta_{\mathrm{R}} ,\varphi_{\mathrm{R}})$ and similarly, $\left\{ b_{\mathrm L},c_{ \mathrm L}\right\}$ to $\left\{ b_{\mathrm R},c_{\mathrm R}\right\}$ across the DW. 
The full solution to $H_{DW}$ is obtained by solving the ordinary differential equation in Eq.~\eqref{eq:EOM} with appropriate boundary conditions (see App.~\ref{sup:AdjBdy} for details) to obtain the tunneling coefficients. 

Let us first consider the uniform eigen-equation $H_{0}(\theta,\varphi) \Psi= \epsilon \Psi$, before placing the DW. A generic solution takes the form
\begin{align}
\Psi_{\bs k} (\bs r) = \mathcal F_s(\eta)\,  e^{i \bs k \cdot \bs r} \quad , \quad
\mathcal F_s (\eta) = \frac{1}{\sqrt 2} \bem 1 \\ s e^{i \eta} \eem .
\label{eq:WaveFun}
\end{align}%
Here $\Psi_{\bs k}$ is a two-component wavefunction in the pseudo-spin basis where the components correspond to the sublattice $A$ of layer 1 and sublattice $B$ of layer 2. $ s = {\rm sign \,} (\epsilon_{\bs k}) =\pm 1$ corresponds to the band index. Without loss of generality we will fix the band index to $s=+1$ since in our case, unlike in the presence of a potential barrier, the chemical potential never passes through two different bands as one moves across the DW
(thereby excluding the presence of any \textit{p-n} or \textit{n-p} junction). Henceforth, we also denote $\mathcal F_+ \equiv \mathcal F$. 

The dependence of $\Psi_{\bs k}$ on $\theta,\varphi$ enters through the phase difference between the two components, $\eta \equiv \text{Arg}(k^{2}-\Delta K^{2})$, which can be obtained by using Eq.~\eqref{eq:WaveFun} in Eq.~\eqref{eq:EOM}. It is worth noting here that the wavefunction $\Psi_{\bs k} (\bs r)$ formally resembles the wavefunction of the low energy electrons in MLG and BLG; except, the phase difference $\eta = \phi_k$ for MLG and $\eta = 2\phi_k$ for BLG, where $\tan \phi_k = k_y/ k_x$ is the angle of propagation. Thus, after a full rotation around a Dirac point, $\Psi_{\bs k}(\bs r)$ obtains a Berry phase of $\pi$ and $2\pi$ for MLG and BLG, respectively~\cite{McCannFalko}. This plays a crucial role in understanding tunneling characteristics of electrons in graphitic systems~\cite{AndoAntiLoc, KleinEPJB11}.  In fact, as we will see, for low energy moir\'e electrons (the ones near the CNP) the phase difference simplifies to $\eta = \varphi + \pi/2 + \phi_q$, which gives rise to a Berry phase of $\pi$ for a closed orbit. This will be useful in understanding the low-energy scattering discussed in App.~\ref{sec:Snell}.

We now compute the transmission and reflection coefficients for a single DW and an array of DWs (along the $x$-direction). Due to the translation symmetry along the $y$-axis, $k_y$ is still a good quantum number, however $k_x$ value switches from $k_{\mathrm{L} }$ on the left to $k_{\mathrm{R} }$ on the right of the DW. The incidence angle is thus, $\tan \phi_{\mathrm{L} } = k_y/ k_{\mathrm{L} }$, while the outgoing angle is $\tan \phi_{\mathrm{R} } = k_y/k_{\mathrm{R} }$. The most general ($L^2$-normalizable) solution for Eq.~\eqref{eq:EOM} can be written in terms of the wavefunction in Eq.~\eqref{eq:WaveFun} as
\bse \label{eq:FullSol}
\begin{align}
x < 0: \quad \Psi_{\mathrm{L} }(\bs r) =& \, \Big[ 
{p_{\mathrm{L} }^+} \, \mathcal F(\eta^+_{\mathrm{L} }) e^{i x k_{\mathrm{L} }} + p_{\mathrm{L} }^- \, \mathcal F(\eta_{\mathrm{L} }^-) e^{- i x k_{\mathrm{L} }} 
\nonumber \\ 
& + e_{\mathrm{L} } \, \mathcal F( - i \log \chi_+)  e^{x \kappa_{\mathrm{L} } } \Big] e^{ i y k_y}   ,
\\
x \geq 0: \quad \Psi_{\mathrm{R} }(\bs r)  =& \, \Big[ 
  p_{\mathrm{R} }^+ \, \mathcal F (\eta_{\mathrm{R} }^+) e^{i x k_{\mathrm{R} } } +p_{\mathrm{R} }^- \, \mathcal F (\eta_{\mathrm{R} }^- )  e^{- i x k_{\mathrm{R} } } 
  \nonumber \\ 
& + e_{\mathrm{R} } \, \mathcal F(- i \log \chi_-) e^{- x \kappa_{\mathrm{R} }} \Big] e^{ i y k_y} .
\end{align}
\ese
Here, $p_j^+$ ($p_j^-$) correspond to the amplitudes of the propagating modes moving to the right (left) on the $j=L,R$ side of the DW. Unlike in MLG, there always exists a pair of exponentially decaying solutions in BLG~\cite{NovoselovKlein} which is why we include the evanescent modes of amplitude $e_{ \mathrm{L,R} }$. For the case of a single domain, assuming no incidence from the right, we would set $p_{\mathrm{R} }^-$ to zero. Thus, $p_{\mathrm{L} }^+$ can also be be normalized to one. The phase difference between the two pseudo-spins can be obtained by solving the eigenvalues of $H_0$,
\begin{align}
\begin{split}
\eta^\pm_j &= {\rm Arg}\left[ (\pm k_j + i k_y)^2 - \Delta K_{\theta_j}^2 \right] 
, \\
\chi_\pm &= - {\rm sgn} \left[ (\pm \kappa_{ \mathrm{L,R} } + k_y)^2 + \Delta K_{\theta_{ \mathrm{L,R} } }^2 \right]  .
\label{eq:Phases} 
\end{split}
\end{align}%
Here ${\rm {Arg}}(z)$ is the principal valued argument of a complex number $z$, and ${\rm{sgn}}
(z) = z/|z|$. When $\varphi_{ \mathrm{L,R} }=0$, one obtains $\chi_\pm=-1$ and $\eta^\pm_j = - \eta_j^\mp \equiv \eta_j$. The wave vectors  corresponding to all the modes can be obtained by solving the characteristic equation of Eq.~\eqref{eq:EOM}. Being a fourth order equation it admits four solutions. The real solutions, $ k_x = \pm k_{ \mathrm{L,R} }$, correspond to the momentum of the propagating modes and the imaginary solutions, $k_x = \pm i \kappa_{ \mathrm{L,R} }$, correspond to the wave vectors of the evanescent modes. Instead of writing their cumbersome general solution, we explicate their dependence on ($\theta, \varphi, k_y$) in relevant sections.

In order to solve the amplitudes of the various modes in Eq.~\eqref{eq:FullSol}, we impose the matching conditions (at $x=0$) obtained in Eq.~\eqref{eq:BCfinal1},
\bse \label{eq:Match}
\begin{align}
\Psi_{\mathrm{L} }(0, y) &= \xi \Psi_{\mathrm{R} }(0, y) , \label{eq:Match1} \\
\partial_x\Psi_{\mathrm{L} }(0, y) &= \xi\partial_x \Psi_{\mathrm{R} }(0, y) + \zeta \Psi_{\mathrm{R} }(0, y)  , 
\end{align}
\ese
where $\xi = a_{\mathrm{L} }/ a_{\mathrm{R} }$ and $\zeta=i \xi \left( b_{\mathrm{R} }/a_{\mathrm{R} } - b_{\mathrm{L} }/a_{\mathrm{L} } \right)/2$. Using these equations we eliminate the amplitudes of the evanescent modes and  obtain the transfer matrix, $\mathcal M$, from
\begin{gather}
\bem p^+_{\mathrm{R} } \\ p^-_{\mathrm{R} } \eem = \mathcal M \bem p^+_{\mathrm{L} } \\ p^-_{\mathrm{L} } \eem .
\label{eq:TransMat}
\end{gather}%
The full form of $\mathcal M$ is obtained in the Appendix~\ref{sup:DeriveM}. Most of our discussion will concern the simple case of $\varphi_{ \mathrm{L,R} }=0$, for which, as mentioned before, $\chi_\pm = -1$, $\eta_j^\pm = - \eta_j^\mp \equiv \eta_j$ and $ \zeta=0$. The normalized transfer matrix takes the form
\begin{align}
\mathcal M = \mathcal M_0
\begin{bmatrix}
\frac{1 + e^{i \eta_{\mathrm{L} }}}{1 + e^{i \eta_{\mathrm{R} }}} \, \left( 1 + \frac{k_{\mathrm{L} }}{k_{\mathrm{R} }} \right) &
\frac{1 + e^{- i \eta_{\mathrm{L} }}}{1 + e^{i \eta_{\mathrm{R} }}} \left( 1 - \frac{k_{\mathrm{L} }}{k_{\mathrm{R} }} \right) \\
\frac{1 + e^{i \eta_{\mathrm{L} }}}{1 + e^{-i \eta_{\mathrm{R} }}} \left( 1 - \frac{k_{\mathrm{L} }}{k_{\mathrm{R} }} \right) &
\frac{1 + e^{-i \eta_{\mathrm{L} }}}{1 + e^{-i \eta_{\mathrm{R} }}} \left( 1 + \frac{k_{\mathrm{L} }}{k_{\mathrm{R} }} \right) 
\end{bmatrix} . 
\end{align}
Here $\mathcal M_0$ is the normalization constant that is fixed by requiring $|\det \mathcal M |^2=1$. From this we obtain the reflection ($R$) and tunneling coefficients ($T$) as
\begin{align}
R &=  \left \rvert \frac{\mathcal M_{21}}{\mathcal M_{22} } \right \rvert^2 = \frac{(k_{\mathrm{L} } - k_{\mathrm{R} })^2}{(k_{\mathrm{L} } + k_{\mathrm{R} })^2} 
\label{eq:Rdef} , \\
T &=  \left \rvert \frac{1}{\mathcal M_{22} } \right \rvert^2 = \frac{4 k_{\mathrm{L} } k_{\mathrm{R} }}{(k_{\mathrm{L} } + k_{\mathrm{R} })^2}  = 1-R.
\label{eq:Tdef}
\end{align}%

It must be noted here that the above expressions are similar to those for tunneling across a step-potential, the reason for which will be evident from the section below. Also, note that the tunneling expression above is real even when there are only evanescent modes (imaginary $k_{ \mathrm{L,R} }$) on either sides of the DW. One must reject these spurious solutions, which arise simply due to the quartic nature of the squared-dispersion. 

\begin{figure}[ht!]
\centering
\quad \includegraphics[width=0.48 \textwidth]{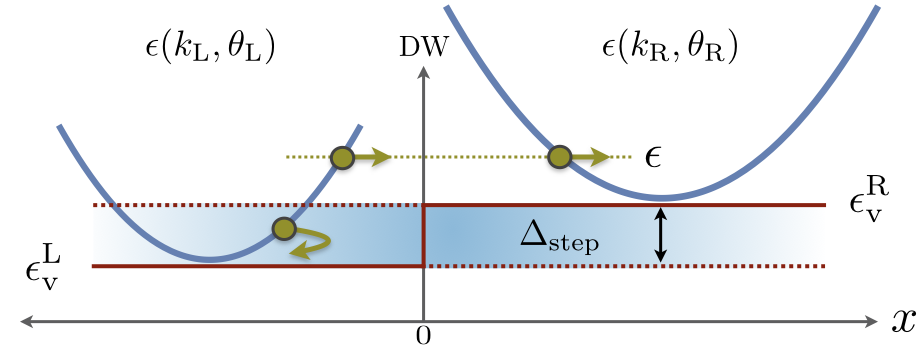}\\ \vspace{2mm}
\includegraphics[width=0.45 \textwidth]{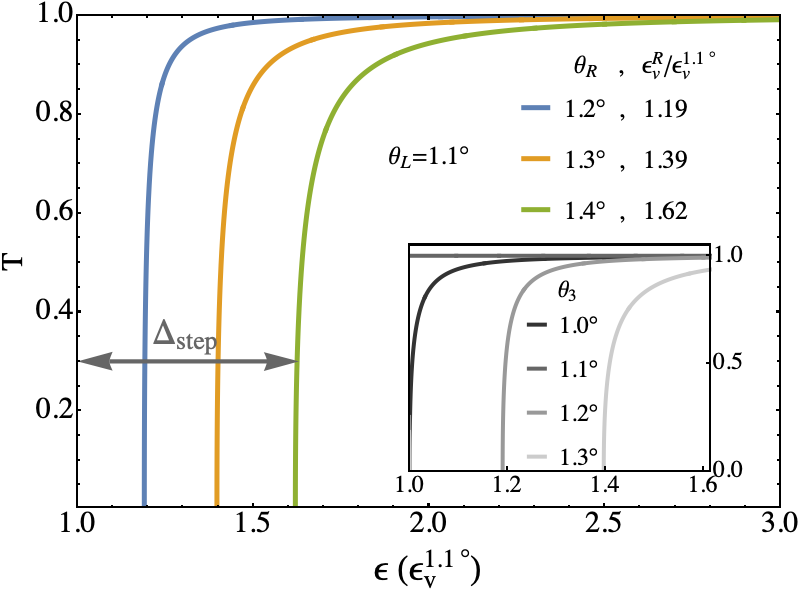} 
\caption{Normal tunneling: 
(Top panel) Electronic dispersions for $k_y=0$ and $\varphi_{ \mathrm{L,R} }=0$ with  $\theta_{\mathrm{L} } < \theta_{\mathrm{R} }$. An electrons (green dot) can tunnel across the DW only when the Fermi level ($\epsilon_F$) is larger than the maximum of the energy minima, ${\rm max} \left( \epsilon_\mathrm{v}^{ \mathrm{L} }, \epsilon_\mathrm{v}^{ \mathrm{R} } \right)$. The blue region delineates the classically forbidden ($R=1$) region, though this can contribute to tunneling if another TBG with $\theta < \theta_{\mathrm{R} }$ is placed to its right (see inset in the lower panel). The gray region does not allow any propagating modes to exist. The tunneling problem considered here can be understood through formation of an effective step-potential of height, $\Delta_{\mathrm{step} } =| \epsilon_\mathrm{v}^{ \mathrm{R} } - \epsilon_\mathrm{v}^{ \mathrm{L} }|$.
(Bottom panel) Normal tunneling across a DW as a function of chemical potential, with $\theta_{\mathrm{L} } = 1.1^\circ < \theta_{\mathrm{R} }$ (see legends) and $\varphi_{ \mathrm{L,R} }=0$. For reasons discussed above, $T$ is finite only when $\epsilon > \epsilon_\mathrm{v}^{1.1^\circ} + \Delta_{\mathrm{step} }$ (legends). Clearly, the larger the difference in twist angles, the larger is the tunneling gap, $\Delta_{\mathrm{step} }$.
(Inset) We place another DW parallel but far away from the first one. The twist angles from left to right are $\theta_1 = 1.1^\circ, \theta_2 = 1.2^\circ, \theta_3$ (see legend). When $\theta_3 > \theta_2$ the effective $\Delta_{\mathrm{step} }$ increases otherwise it decreases. For the case $\theta_1=\theta_3$ resonant tunneling occurs since the entire blue region is now allowed to tunnel, see Sec.~\ref{sec:DWarray} for details.
}
\label{fig:NormTun}
\end{figure}

Though not manifest, $T$ and $R$ are dependent on $\theta, \varphi$ through the $x$-axis momenta, $k_{\mathrm{L,R }}$. We analyze this dependence, first, in case of a longitudinal DW and then for a tilted DW -- when there is a finite tilt angle between the DW and the $K$-$K'$ zone boundary of the MBZ.

\section{Longitudinal Domain Walls}
\label{sec:LongiDW}

For $\varphi = 0$ the wave vectors are obtained to be
\begin{align}
\begin{split}
k_j^2 &= - k_y^2 - K_{\theta_j}^2 + \sqrt{\epsilon^2  + 4 k_y^2 K_{\theta_j}^2 }
\label{eq:kxeqn}
\\
\kappa_j^2 &=  k_y^2 + K_{\theta_j}^2 + \sqrt{\epsilon^2  + 4 k_y^2 K_{\theta_j}^2 } ,
\end{split}
\end{align}%
the spinor phases simplify to $\eta^\pm_j = \eta_j$ and  $\chi_\pm =-1$. In this section, we analyze tunneling of gapped (high-energy) moir\'e states, which are closer to the $M$ point or of energy of the order of $\epsilon_\mathrm{v}$. We postpone our discussion on tunneling of the (low-energy) moir\'e states near the CNP to the subsequent section, since, as we will see, their tunneling characteristics are independent of the value of the tilt angle. 
We first focus on normal tunneling, then we discuss oblique tunneling. Independent of the tilt angle, as can be seen from Eq.~\eqref{eq:Dispersion}, normally incident quasiparticles are maximally gapped ($= 2 \epsilon_\mathrm{v}$). For particles incident obliquely, this gap reduces. We will see this effective gap for a given $\theta_j, k_y$, play a crucial role in controlling tunneling across the DW. We conclude the section by extending our discussion on tunneling in presence of multiple DWs and in case of a smooth DW.

\subsection{Normal Incidence: Step Potential}
\label{sec:Normal}

In this subsection we analyze the tunneling expression in Eq.~\eqref{eq:Tdef} for electrons near the $M$ point for normal tunneling. Strictly speaking, normal incidence refers to vanishing (transverse) group velocity, $v_y = \nabla_{k_y} \epsilon_k = 0$. For our dispersion this occurs for $k_y=0$ and $|k| = K_{\theta_j}$, which traces a circle centered at the $M$ point traversing through the two neighboring Dirac points. For brevity, we will refer to $k_y=0$ only as ``normal tunneling'' and we treat the other case as an instance of oblique tunneling in Sec.~\ref{sec:Oblique}.

Using Eq.~\eqref{eq:kxeqn} in Eq.~\eqref{eq:Tdef} we obtain the tunneling probability for normal incidence, which is plotted in the bottom panel of Fig.~\ref{fig:NormTun}. The switching behavior of $T$ can be understood from its top panel, which shows the electronic dispersion across the DW, in particular, for the case $\theta_{\mathrm{R} } > \theta_{\mathrm{L} }$. For energy in the gray region, $0 < \epsilon < \epsilon_\mathrm{v}^{ \mathrm{L} }$, there are no propagating modes on either side of the DW, thus tunneling is prohibited. Here $\epsilon_\mathrm{v}^{\mathrm{L},\mathrm{R}}$ are the band minima of the left and right sides, respectively. For $\epsilon > \epsilon_\mathrm{v}^{ \mathrm{R} }$, for every propagating state on the left there is a propagating state available on the right, thus tunneling is perfect ($T \simeq 1$). However, in the blue region, $\epsilon_\mathrm{v}^{ \mathrm{L} } < \epsilon < \epsilon_\mathrm{v}^{ \mathrm{R} }$, just the left side has a propagating mode, hence it can only contribute to (perfect) reflectivity. This difference in the vHS energies across the DW, $\Delta_{\mathrm{step} } = |\epsilon_\mathrm{v}^{ \mathrm{L} } - \epsilon_\mathrm{v}^{ \mathrm{R} } |$, is precisely what offers an effective realization of a step-potential [see Eq~\eqref{eq:Tdef}] of height $\Delta_{\mathrm{step} }$. This naturally manifests as a gap in tunneling, which grows with increasing twist angle difference. As $\epsilon$ overcomes this gap, tunneling rapidly switches to one. 

It must be noted that, for the case of $\theta_{\mathrm{R} } < \theta_{\mathrm{L} }$ the step potential essentially switches to a down-hill potential. Since incidence still remains from the left, as long as $\epsilon > \epsilon_\mathrm{v}^{ \mathrm{R} }$ there will always be tunneling. 
In order to obtain tunneling behavior for this case one simply needs to interchange $k_{\mathrm{L} }$ and $k_{\mathrm{R} }$. This yields tunneling curves similar to those in Fig.~\ref{fig:NormTun} except reflected around the $\epsilon = \epsilon_\mathrm{v}^{ \mathrm{L} } $ line. 

\subsection{Oblique Incidence: Reduced Step Size }
\label{sec:Oblique}

\begin{figure}[t!]
\includegraphics[width= 0.47 \textwidth]{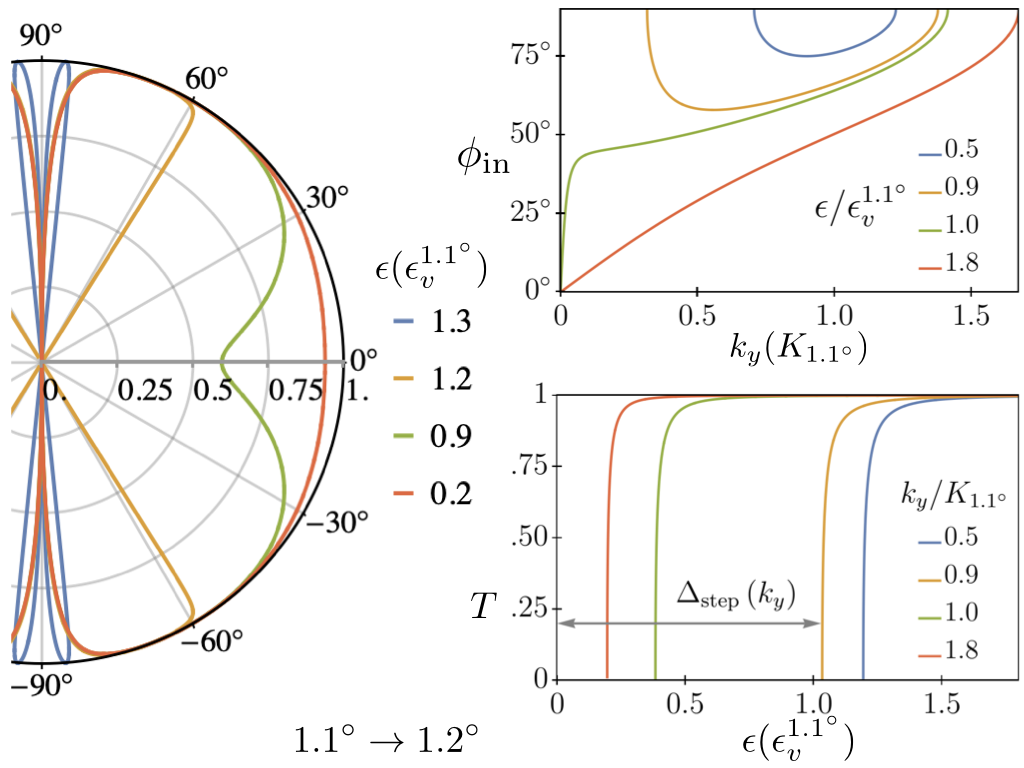}  
\caption{Oblique Incidence for $(\theta_\mathrm{L}, \theta_{\mathrm{R}}) = (1.1^\circ, 1.2^\circ)$. The left panel shows a polar plot of tunneling amplitude as a function of quasiparticles incidence angle. The curves are obtained by numerically solving Eq.~\eqref{eq:Tdef} for various energies (measured in the units of $\epsilon_\mathrm{v}^{ \mathrm{L} }$). All four quadrants in the polar plot are symmetric. The polar spread of the curves decreases with decreasing energy. This is explained through the top right panel. The allowed value of incidence angle or $k_y$ momentum for propagating states decreases as one approaches the Dirac point. Note, for a given value of $\phi$, there can be more that two real propagating modes, especially for $\epsilon \lesssim \epsilon_\mathrm{v}$. (Bottom right) plotting the tunneling as a function of energy demonstrates the $k_y$ dependence of the $\Delta_{\mathrm{step} }$.
}
\label{fig:Oblique}
\end{figure}

We show the numerically obtained tunneling probability for oblique incidence in Fig.~\ref{fig:Oblique}. Note that, first of all,  for a given energy, the value of $k_y$ (hence, that of the incidence angle) cannot be arbitrarily large. This can be understood from the following relation
\begin{align}
k_y^2 = \sin^2 \phi \left[ - \cos 2 \phi \pm \left( \epsilon^2 - \sin^2 2 \phi \right)^{1/2} 
\right] .
\end{align}%
In fact, for a finite $k_y$, tunneling can be finite only when $\epsilon > \epsilon_\mathrm{v}^{ \mathrm{R} } - k_y^2 $. Thus, the larger the $k_y$, for a given set of $\theta_{ \mathrm{L,R} }$, the smaller is the energy required to tunnel. In other words, since an obliquely incident moir\'e particle experiences a reduced $\Delta_{\mathrm{step} }$, its tunneling is enhanced as compared to that of the normally incident particle. 
Secondly, the tunneling at $\phi=0$ (normal incidence) is strongly dependent on the value of incident energy, in fact it can range anywhere from $0$ to $1$ by suitably adjusting the energy. This is markedly different~\cite{NovoselovKlein} from the case of MLG (where it is always 1), or from BLG~\cite{MassiveChiral} (where it is always zero), and thus a unique characteristic of TBG based barriers.

\subsection{ Array of DWs: Resonant Tunneling}
\label{sec:DWarray}

Now let us consider tunneling across two consecutive DWs separating three TADs of twist angles $\theta_1, \theta_2, \theta_3,$ ordered from the left to the right. Accordingly, we denote the incident $x$-axis momenta as $k_{1,2,3}$ and the spinor phases as  $\eta_{1,2,3}$. For simplicity, we fix the tilt angles in all three domains to be zero. We show that the quantum tunneling in the (blue) classically forbidden region discussed in Fig.~\ref{fig:NormTun} can now resume with the help of the evanescent modes in the middle domain. 

By assuming the DWs are significantly far away from each other on a moir\'e lattice scale, we apply the transfer matrix formalism discussed previously to obtain
\begin{align}
T = &\; 1 - \frac{\mathcal K - 4 k_1 k_2^2 k_3 }{\mathcal K + 4 k_1 k_2^2 k_3  }, \nonumber \\
\mathcal K =&\;  (k_1^2 + k_2^2) (k_2^2 + k_3^2) \nonumber \\
&\;  + (k_1^2 - k_2^2) (k_2^2 - k_3^2) \cos \left(\eta_1 - \eta_3 \right) .
\end{align}
The phase mismatch term, $\cos \left(\eta_1 - \eta_3 \right)$, between the first and the last domains plays a crucial role. When this is equal to identity, tunneling simplifies to 
\begin{align}
T \big \rvert_{\eta_1 = \eta_3} = \frac{ 4 k_1 k_3 }{(k_1 + k_3)^2} .
\end{align}%
Clearly, this does not depend on $k_2$ and thus is independent of the twist angle of the intermediate domain. There are two important scenarios in which this can be achieved. For normal tunneling, since $\eta_1 = \eta_3=0$, see Eq.~\eqref{eq:Phases}. Also, when the twist angle of the leftmost and rightmost regions are the same, irrespective of $\theta_2$ and $k_y$, we have, following Eqs.~\eqref{eq:Phases} and~\eqref{eq:kxeqn}, $k_1 = k_3$ and $\eta_1 = \eta_3$. In fact, this reduces the above tunneling expression to $T=1$, an instance of resonant tunneling. This is depicted in the inset of Fig.~\ref{fig:NormTun}. Such resonant tunneling occurs since for $\theta_1 = \theta_3$ [hence, $\epsilon_\mathrm{v}^{(1)} = \epsilon_\mathrm{v}^{(3)}$] evanescent modes corresponding to any energy (the entire blue region in Fig.~\ref{fig:NormTun}) participate in tunneling. For DWs of finite width, $T=1$ may receive some correction. 

We now generalize the above result to an array of $n$ DWs. Multiplying all the $n$ transfer matrices corresponding to each of the DW, we obtain $T$. When the pseudospin phases of all the domains match, such as for normal incidence, we inductively establish 
\begin{align}
T \big \rvert_{\eta_1 = \eta_2 \cdots = \eta_{n+1} } = \frac{ 4 k_1 k_{n+1} }{(k_1 + k_{n+1})^2} .
\end{align}
Thus, tunneling of normally incident moir\'e electrons is decided \textit{only} by the first and the last twist angles. Although this result is obtained for the case of an array of longitudinal DWs, one can generalize this to any orientation of the MBZ. In Sec.~\ref{sec:Tilted} we will show this for a tilted MBZ, that is, tunneling of normally incident electrons is marginally affected by the tilt angle. Thus, irrespective of the orientation of the MBZ in all the domains and independent of the twist angles in all the intermediate domains, the above conclusion remains robust.

\begin{figure}[t]
\centering
\subfloat
[]{\includegraphics[width=0.25 \textwidth]{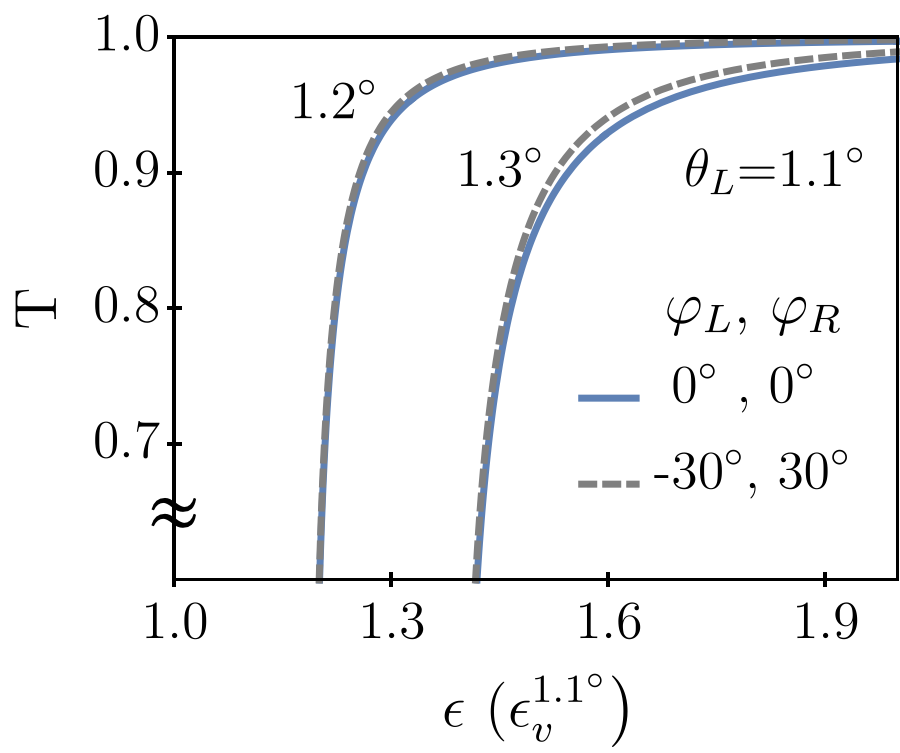} 
\label{fig:FinVarPhi} }  
\subfloat 
[]{ \includegraphics[width=.215 \textwidth]{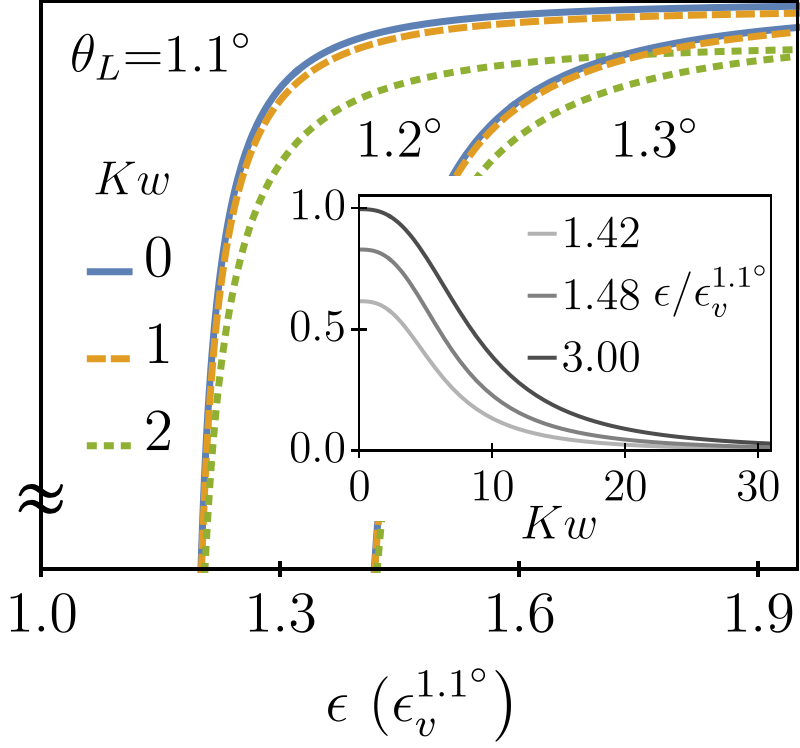} 
\label{fig:FinWidth} }
\caption{
Corrections to normal tunneling: (a) for finite $\varphi_j$, $T$ receives an perturbatively corrected. This result is obtained in Eq.~\eqref{eq:PertPhiT}. The solid (blue) curve is for a longitudinal DW and the dashed (gray) curve is for a tilted DW.
(b) When the DW is considered to be of small but finite width, tunneling receives a decremental correction, see Sec.~\ref{sec:Smooth}. The widths ($w$) of the DWs are indicated in units of inverse Dirac momentum. The inset, drawn for different energy slices of the $(\theta_{\mathrm{L} }, \theta_{\mathrm{R} }) = (1.1^\circ, 1.3^\circ)$ curve, shows that tunneling probability decreases rapidly with increasing width of the DW. 
}
\label{fig:Corrections}
\end{figure}

\subsection{Smooth DW: Exponential Suppression}
\label{sec:Smooth}

The DWs we have considered so far can often be of certain thickness. In other words, two TADs of twist angles $\theta_{ \mathrm{L,R} }$ may contain an intermediate region where $\theta_{\mathrm{L} }$ smoothly, as opposed to abruptly, changes to $\theta_{\mathrm{R} }$. In this subsection we discuss tunneling across such a smooth DW.

Before proceeding to compute tunneling, first we compare the various length scales and their relevance in the problem. For a TBG with $\theta \approx 1^\circ$ the moir\'e periodicity is $\lambda=13\, \text{nm}$, which is about 50 times the lattice constant of pristine graphene, $a_0=0.25\,$nm. Therefore, since $|\bs K - \bs K'| \sim \lambda^{-1} \ll a_0^{-1}$, different valleys of the original graphene layers are decoupled. Within a single valley, the mini-bandwidth is of the order of $10$ meV~\cite{KimReconstruction,MCDMoire,NamKoshino17}. Hence, the Fermi wavelength of the moir\'e electrons is at least, $\lambda_{\mathrm{F}} \gtrsim 12\,$nm. This means Umklapp processes involving inter-mini-valley scatterings should be feasible~\cite{AshvinValley,ValleyJT}. In order to exclude such processes in out study, we avoid getting closer to the zone center of the MBZ where $\lambda_F$ is the smallest, i.e., $\sim \lambda$. 

When a DW is of width $w$ and $w \ll  \lambda_F$ then it is sufficient to treat the DW as a sharp boundary. This is especially more reasonable for low-energy quasi-particles, say near  charge neutrality (unless one fabricates a junction of several $a_0$ between two TBG devices). On the other hand, as the quasiparticle energy increase, or $\lambda_F$ gets shorter, one must take the finite width of the DW into account. In a recent experiment~\cite{uri2019mapping} a spatially varying $\theta(\bs r)$ was observed to have $\partial_{x} \theta = (0.02^\circ - 0.05^\circ)/ \mu$m (that is about $2\% - 5\%/ \mu$m). This reflects a 1\% change in twist angle in about every $10-15$ moir\'e periods.
Thus, unless one is near charge neutrality, it is worth considering the perturbation for wide DWs. For simplicity we will do so near the $M$ point by setting $k_y=0$.

Consider a DW of width, $w \gg \lambda$, over which the twist angle between the two regions changes smoothly, $\theta(x)$. Again, this is smooth in the scale of $\lambda$ or $K_\theta^{-1}$. In the limit of a slowly varying $\theta(x)$ we can still work with our original Hamiltonian, $H_0$, and define $K_{\theta(x)} \simeq \theta(x)$. We start by writing the equation of motion for the two sublattice wavefunctions as
\bse
\begin{align}
\partial_x^2  \psi_B  - K_{\theta(x)}^2  \psi_B + \epsilon \psi_A = 0 , \\
\partial_x^2  \psi_A  - K_{\theta(x)}^2  \psi_A + \epsilon \psi_B = 0 .
\end{align}%
\ese
Like before, we will be interchangeably using $K_\theta(x)^2$ and $\epsilon_\mathrm{v}(x)$, a position dependent {band minimum}. The above system of equations can be decoupled by rotating the basis to $\psi_\pm = \left( \psi_A \pm \psi_B \right)/\sqrt{2}$ and have
\begin{align}
\left[ - \partial_x^2  + \epsilon_\mathrm{v}(x) \right] \psi_\pm  = \pm \epsilon  \psi_\pm .
\label{eq:SmoothEOM}
\end{align}
This is similar to the equation of motion for a Schr\"odinger particle (or hole) in a potential generated by $\epsilon_\mathrm{v}(x)$. We can define an effective momentum for such particles as $p^2_\pm(x) = \pm \epsilon - \epsilon_\mathrm{v}(x)$, and write the above equation as $-\partial^2_x \psi_\pm = p_\pm^2 \psi_\pm$. Clearly, $p_-^2$ is always negative for any positive value of $\epsilon$. Thus, being an evanescent mode, $\psi_-$ never contributes to tunneling. The same is true for $\psi_+$ unless $\epsilon > {\rm max} [\epsilon_\mathrm{v}(x)]$. For a generic profile of $\theta(x)$ one could always solve the above equation in the semiclassical limit,
\begin{align}
\psi_+(x) \simeq \frac{c_\pm}{\sqrt{2p_+(x) }} \, e^{\pm \gamma(x)} \,\, , \,\, \gamma(x) = \int_{0}^x p_+(y) dy \, .
\end{align}%
For instance, one could take the experimental $\theta(x)$ and convert it into the $p_\pm(x)$ above. The tunneling probability can then be obtained using the Wentzel--Kramers--Brillouin (WKB) approximation~\cite{FalkoPRB06, GHG-smoothPN}
\begin{align}
T \simeq e^{-2 \gamma(w)} \, .
\end{align}%
There are two things to be noted here -- firstly, the upper limit of integration above is $w$, which we have assumed to be a turning point. Otherwise, for a turning point at $x<w$, $\epsilon $ would have to satisfy $\epsilon < \epsilon_\mathrm{v}(x)$, in which case there would not be any tunneling. Second, the WKB expression above is valid only when $\gamma$ itself is large (or $T \ll 1$), in other words, this requires the domain to be very wide, $w k_F \gg 1$. Thus, the above expression is applicable strictly when $\epsilon$ starts overcoming $\Delta_{\mathrm{step} }$ and tunneling slowly ramps up from zero. The wider a domain, the larger is $\gamma$ and the smaller is $T$, or it increases with a much slower rate.

We demonstrate the above conclusions by explicitly (and exactly) evaluating tunneling for a linearly changing $\theta(x)$. In fact, for simplicity, we approximate the corresponding vHS energy to be
\begin{align}
\epsilon_\mathrm{v}(x) \simeq K^2 \bar \theta \left( \bar \theta + 2 x \nabla \theta \right)/4 \quad , \quad \theta(x) = \bar \theta + x \, \nabla \theta,
\end{align}%
with $\left( \bar \theta,  w \nabla \theta \right) = \frac12 \left(\theta_{\mathrm{R} } \pm \theta_{\mathrm{L} }\right) $. For such a linear profile we can solve the eigen-equation~\eqref{eq:SmoothEOM} exactly, similar to the case of a triangular potential well,
\begin{gather}
\psi_\pm (x) = \alpha_\pm \, \Ai{z_\pm (x)} + \beta_\pm \, \Bi{z_\pm (x)} ,  \nonumber \\
 z_\pm(x) = \frac{\epsilon_\mathrm{v}(x) \mp \epsilon}{(K^2 \bar \theta \nabla \theta/2)^{2/3} } .
\end{gather}
Here, $\Ai{z}$ and $\Bi{z}$ are the Airy functions of first and second kind, respectively. Note, equations of motion corresponding to $\theta(x)$ as higher order polynomials can also be solved similarly using parabolic cylinder functions, $D_\nu(z)$,~\cite{Abra72}. With the wavefunctions above, we repeat the transfer matrix method discussed in Sec.~\ref{sec:Tunneling} and obtain the tunneling probability. The results are plotted in  Fig.~\ref{fig:FinWidth}. Indeed, as the width of the DW increases, $w K \rightarrow \infty$, it becomes exponentially harder for the moir\'e electrons to tunnel across two TADs. Similar suppression due to widening of an otherwise sharp potential step is also seen for Klein tunneling~\cite{FalkoPRB06}.

\section{Tilted Domain Walls} 
\label{sec:Tilted}

In this section we study tunneling across a DW that is at a finite angle with respect to the zone boundary of the MBZ. In particular, we study the effect of a finite $\varphi$ on normal tunneling near the $M$ point. Although the exact treatment is rather tedious, we can perturbatively understand its effects for the desired range of $\varphi$ which may be as large as $30^\circ$. 

For normal incidence we have, 
\begin{align}
k^2_x = - K_\theta^2 \cos 2\varphi \pm \sqrt{\epsilon^2 - \epsilon_\mathrm{v}^2 \sin^2 2\varphi} \,,
\label{eq:normalkx}
\end{align} 
which is always real for $\epsilon > \epsilon_\mathrm{v}$. Therefore, for this energy window, there exists a pair of propagating modes which \textit{may} contribute to tunneling. Setting $k_{y}=0$ while keeping the tilt angle $\varphi_j$ finite, we expand Eq.~\eqref{eq:Phases} and Eq. \eqref{eq:normalkx} up to $\mathcal O \left(\varphi^{4}_{j} \right)$ and obtain
\begin{gather}
\chi_j \simeq \; - 1 + 2 i \varphi_j \tilde \epsilon_j +2 \varphi_j^2 \tilde \epsilon_j^2 \,\, , \,\,
e^{i\eta^{\pm}}\simeq\; 1 + 2i\tilde{\epsilon}_j\varphi_j - 2\tilde{\epsilon}_j \varphi_{j}^{2} ;
 \nonumber \\
k_{j}\simeq \; k_{j}^{0}(1+\tilde{\epsilon}_{j}\varphi^{2}) \quad , \quad
\kappa_{j}\simeq \; \kappa_{j}^{0}(1-\tilde{\epsilon}_{j}\varphi_{j}^{2}) .
\end{gather}
Here, $\tilde \epsilon_j \equiv \epsilon_\mathrm{v}^j/ \epsilon$, $k_j^0 \equiv \left( \epsilon - \epsilon_\mathrm{v}^j \right)^{1/2} $ and $\kappa_j^0 \equiv \left( \epsilon + \epsilon_\mathrm{v}^j \right)^{1/2} $. We note here that at order $\mathcal O \left(\varphi^{4}_{j} \right)$, the above expressions are fairly accurate for $|\varphi_j | \lesssim 30^\circ$. In fact, due to the 6-fold symmetry of the MBZ it is sufficient to consider $| \varphi_j | \leq 30^\circ$.

Using the above expressions we re-evaluate the transfer matrix in Eq.~\eqref{eq:TransMat}. This obtains the tunneling probability to be
\begin{align}
\frac{T}{ T^{(0)} } \simeq&\;  \, 1 
+ \sqrt{R^{(0)}} \Big[  \varphi_{\mathrm{R} }^2 \tilde \epsilon_{\mathrm{R} } - \varphi_{\mathrm{L} }^2 \tilde \epsilon_{\mathrm{L} }  \nonumber \\
&\; +  \frac{\kappa_{\mathrm{L} }^0 - \kappa_{\mathrm{R} }^0}{\kappa_{\mathrm{L} }^0 + \kappa_{\mathrm{R} }^0}  \left(\varphi_{\mathrm{L} } \tilde \epsilon_{\mathrm{L} } - \varphi_{\mathrm{R} } \tilde \epsilon_{\mathrm{R} } \right)^2 \Big] + \mathcal{O}(\varphi_j^3).
\label{eq:PertPhiT}
\end{align}%
Here, $R^{(0)}$ and $T^{(0)}$ are the reflection and tunneling coefficients for the $\varphi=0$ case, as obtained in Eqs.~\eqref{eq:Rdef} and~\eqref{eq:Tdef} using $k_j^0$, respectively. We note the following about the above expression. Firstly, the correction term is proportional to $T^{(0)}\sqrt{R^{(0)}}$. Recall, $R^{(0)}$ quickly vanishes as the Fermi level moves above the gap, $\Delta_{\mathrm{step} }$, see Fig.~\ref{fig:NormTun}. And, when the Fermi level is inside the gap $T^{(0)}$ vanishes. Therefore the effect of the correction term can never be significant. This is indeed what we observe, see Fig.~\ref{fig:FinVarPhi}. Secondly, analyzing the terms inside the square bracket above, we note that, though negligible, the evanescent modes also contribute to tunneling in case of a tilted MBZ. Lastly, we note that the correction term is invariant under the operation $(\varphi_{\mathrm{L} }, \varphi_{\mathrm{R} }) \rightarrow (- \varphi_{\mathrm{L} }, -\varphi_{\mathrm{R} })$. In fact, this is the same operation as reflecting either of the MBZs, for fixed $\varphi_j$, about the $x$-axis. For normally incident electrons, this is clearly a symmetry of the theory as can also be seen in Eq.~\eqref{eq:Dispersion}. This explains the absence of any $\mathcal O(\varphi_j)$ correction term in the above expression.

For finite $\varphi_j$, solutions to oblique tunneling could be cumbersome. however, one may qualitatively understand tunneling using similar arguments as before -- tunneling will switch on once the chemical potential crosses the band minimum (depending on $k_y$, see Fig.~\ref{fig:Schema}, and $\varphi_{ \mathrm{L,R} }$) on either sides of the DW.

\section{Transport Across Domain Walls}
\label{sec:Transport}

In this section we compute the mesoscopic conductivity of a TBG device and discuss how it is affected by the presence of a DW. We then compute the Fano factor for shot noise and show how  the peaks in its response can be used for diagnosing twist disorder.

\subsection{Conductivity} 

\begin{figure}[t!]
\centering
\subfloat[]{ \label{fig:Conduct}
\includegraphics[width=0.23 \textwidth]{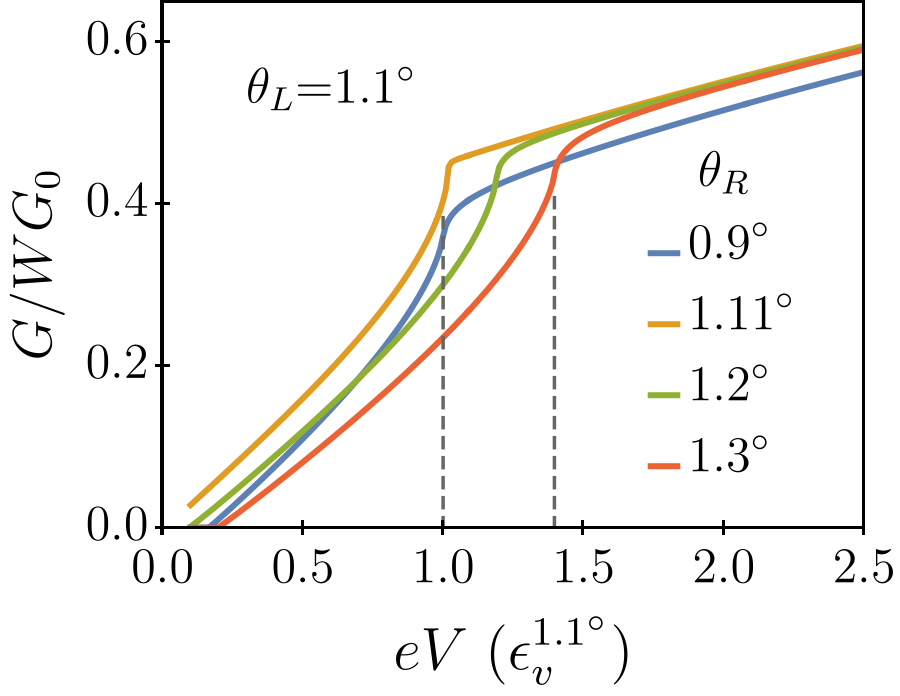} }
\subfloat[]{ \label{fig:Fano}
\includegraphics[width=0.238 \textwidth]{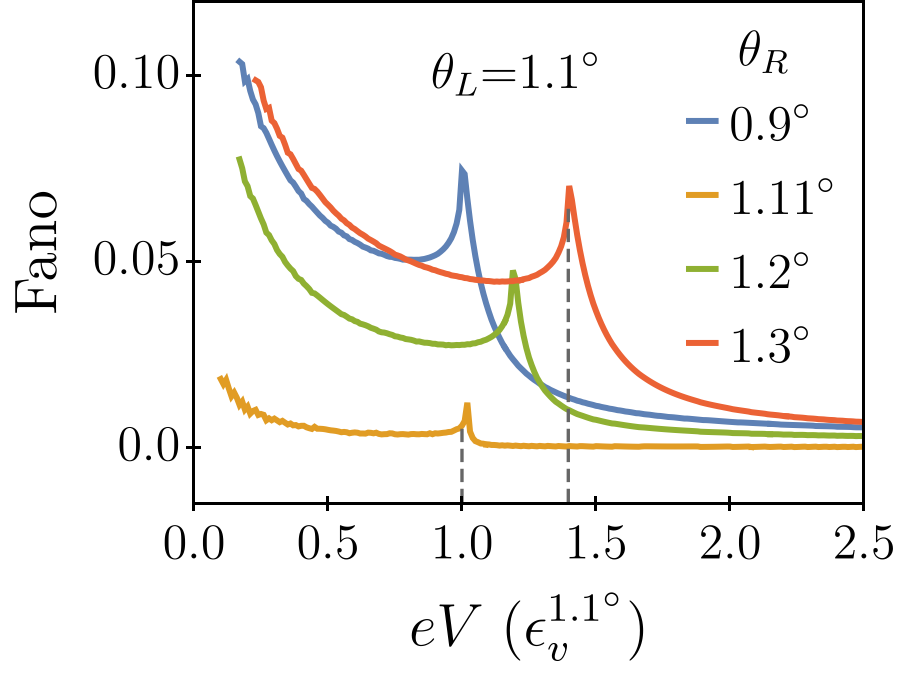}  }
\caption{Quantum transport using Landauer-B\"uttiker formalism: (a) irrespective of the twist disorder, since there are no tunneling states for zero energy, conductance vanishes at zero bias. In presence of twist disorder, a transport gap appears that scales with $\Delta_{\mathrm{step} }$. The location of the inflection point, at $ \epsilon = \epsilon_\mathrm{v}^{ \mathrm{L} } + \Delta_{\mathrm{step} } \equiv \epsilon_{\ast}$, characterizes the strength of the twist disorder. (b) Due to the absence of tunneling, Fano factor at $V=0$ is close to the Poisson value of $1$ (not shown here owing to reduced numerical stability). As tunneling increases, $F$ reduces and ultimately vanishes for high bias potential (since for high energies, $T=1$). A peak appears at $\epsilon_\ast$, the height of which scales with $\Delta_{\mathrm{step} }$.
}
\label{fig:Landauer}
\end{figure}

In low-temperature systems with very few impurities, such as in graphene based materials, the mean free path of the charge carriers can be as large as the size of the sample, giving rise to ballistic transport. In this limit, one can invoke quantum mechanical properties of charge carriers to describe their conduction. In particular, when transport is coherent (single wavefunction extending from one lead to another), the exclusion principle has no effect on conductivity and it can be described using Landauer-B\"uttiker formalism~\cite{SDattaBook}; see also~\cite{KatsnelsonLanduer, BeenakkerNoise, RyuTwistedBC, Guinea-SLG-BLG}. The differential conductance in this mesoscopic limit becomes
\begin{align}
\frac{dI}{dV} =& \, G_0 W  \int  \frac{d}{d\epsilon} f(\epsilon - eV) \, d\epsilon \, \int  \frac{d k_y}{2\pi} \, T(k_y, \epsilon) \nonumber \\
\quad \simeq& \, G_0 W \int   \frac{d k_y}{2\pi} \, T(k_y, e V) \equiv G(V)  .
\label{eq:dIdV}
\end{align}
The last simplification was done taking the zero temperature limit [$V \gg ( e \beta)^{-1}$, with $\beta$ as the inverse temperature] which reduces the Fermi function, $f(\epsilon)$, to a step function. We have also set $\epsilon_{\mathrm{F}}=0$, thereby focusing only on the CNP. $ G_0 = {\rm g} {e^2}/{h} $, where $\rm g = 4$ is a symmetry (valley and spin degeneracies) factor. $W$ is the width of the sample, which, for the applicability of the above formula, should not be much larger than the Fermi wavelength of the moir\'e electrons, $\lambda_F \sim \mathcal O(100 \, {\rm nm})$. In Fig.~\ref{fig:Conduct}, we plot the dimensionless conductivity, $G(V)/WG_0$, as a function of bias voltage $V$ at the CNP.

We note the following features of the above $dI/dV$ characteristic. Much like in a semi-metal~\cite{SMRMP}, conductivity vanishes as the bias voltage goes to zero. In particular, close to zero bias, $G(V)$ vanishes linearly. This is unsurprising since the DOS also vanishes linearly as one approaches the CNP, see Eq.~\eqref{eq:DOSzero} in App.~\ref{sup:DOS}. The corrections from the higher order  terms in the DOS indeed manifest in the $G(V)$ for higher energies. However, with increasing twist disorder a transport gap appears in the $dI/dV$ plot. This gap is a manifestation of the height of the step potential, $\Delta_{\mathrm{step} }$. Therefore, for $\theta_{\mathrm{L} } = 1.1^\circ$, the gap for $\theta_{\mathrm{R} } = 1.3^\circ$ is the largest in Fig.~\ref{fig:Conduct}, whereas that for $\theta_{\mathrm{R} }=1.11^\circ$ is nearly zero. Taking the finite width of the DW into account, or with addition of more TADs, this gap can grow further. A similar transport gap is also observed in tunneling across stacking domains in BLG~\cite{StackingBdys}.

In the low-energy region, for a fixed $\theta_{\mathrm{L} }$ and $V$, the value of $G$ gets smaller  with increasing $\theta_{\mathrm{R} } - \theta_{\mathrm{L} }$. This is reasonable since with increasing disorder or $\Delta_{\mathrm{step} }$ (recall the effective step potential picture) tunneling gets suppressed.
Eventually, for $\epsilon \gg \epsilon_\mathrm{v}$, the curves collapse to a linear plot with a much smaller slope. Clearly, this `inflection point' is itself the saddle point energy, $\epsilon_\mathrm{v}/\epsilon_\mathrm{v}^{1.1^\circ}$, hence dependent on the strength of twist disorder. The collapsing of plots is expected since the DOS for high-energy moir\'e electrons is independent of the twist angle, see Eq.~\eqref{eq:DOSconstant}.

When comparing the above $G(V)$ with experiments, one needs to be careful about two things -- first, transport in TBG near magic angle is dominated by strong correlation. Hence, our non-interacting conductivity may not match well with the experimental conductivity of magic angle samples. However, depending on the strength of the twist disorder, the transport gap mentioned above will cause the the V-shaped differential conductance to become a U-shaped curve. Secondly, unlike the case of MLG, conductivity contribution from twist disordered transport does not have a non-vanishing minimum value~\cite{Ziegler, GrapheneBridge}. This is because in MLG, Klein tunneling renders it highly transparent ($T\approx 1$). However, tunneling across TADs is not transparent, especially for low-energy particles. This can be understood by the following argument. Momenta $k_y \approx K_{\theta_{\mathrm{L} }}$ corresponds to particles near the gapless Dirac point on the left domain. However, since $k_y$ is conserved, whenever $\theta_{\mathrm{L} } \neq \theta_{\mathrm{R} }$, this value of $k_y$ would be far away from the Dirac point in the right domain, i.e., at $k_y=K_{\theta_{\mathrm{R} }}$. In fact, as can be seen from the right panel in Fig.~\ref{fig:Schema}, this difference amounts to opening of a gap on the right side. Therefore, for low energies, the gapless particles on the left cannot scatter into the gapped particles on the right [see App.~\ref{sec:Snell} for a discussion on some interesting consequences of when such tunneling is allowed], leading to vanishing tunneling. Hence, the experimentally observed minimum conductance of TBG \textit{cannot} be shifted by the presence of twist disorder, although conventional disorder can do so~\cite{Koshino-Ando, SDS-Hwang-Rossi}.

\subsection{Fano Response} 

In absence of a tunneling barrier, one expects only thermal noise. This is simply a measure of the temperature of the system. However, in systems where current is partitioned, such as in \textit{p-n} junctions~\cite{LevitovFano, KumadaNoise} or in our case, transport is expected to be noisy even at zero temperature [see App.~\ref{sup:Noise} for a pedagogical introduction to noise in quantum transport]. At zero temperature, noise power (per volt) corresponding to low frequency [of the order of $100 \, \text{MHz}$] shot noise is 
\begin{align}
\frac{dS}{dV} = \frac{{\rm g}  e}{h} \int dk_y \, T(k_y, eV) R(k_y, eV)  .
\end{align}%
Combining this with Eq.~\eqref{eq:dIdV} one can obtain the ratio of noise power per mean current, or the differential Fano factor,
\begin{align}
F(V) &= \frac{1}{ 2 e } \frac{dS}{dV} \large / \frac{dI}{dV} = \frac{1}{ 2 e }  \frac{dS}{dI} \nonumber \\
	&= \frac{G_0}{ e {\rm g} G(V) } \int d k_y  \, T(k_y, eV) R(k_y, eV) \,. 
\label{eq:FanoDef}
\end{align}%
Note the differential Fano factor should not be confused with the average Fano factor, $\int_0^I F(i) di / i$. Although for very large voltage, $V \gg ( e \beta)^{-1}$, the above factor does approach the average Fano factor~\cite{NoiseNanotubes}. $F$ provides crucial information regarding the nature of {the dominant} scattering process~\cite{BeenakkerNoise, GHG-smoothPN}. In particular, a large Fano factor, $F \sim 1$, signals noisy and diffusive transport. On the other hand, $F=0$ characterizes a noise-free ballistic transport.

Fano factor is obtained in Fig.~\ref{fig:Fano} by numerically evaluating Eq.~\eqref{eq:FanoDef}. 
Most importantly, sharp peaks appear in the differential Fano factor which are located at energy equal to $\epsilon_\mathrm{v}^{ \mathrm{L} } + \Delta_{\mathrm{step} }$. Note that such sharp peaks may get smeared at finite temperature. Further, the height of the peak is larger for bigger $\Delta_{\mathrm{step} }$. In other words, the larger the twist disorder, noisier is the transport. For very high bias potential, the Fano factor vanishes, because for very high energy the DWs are transparent ($T\approx1$). On the contrary, for very low temperature, since tunneling probability is very low (due to the step potential), transport becomes noisy. Clearly, in the absence of twist disorder all such features are absent and $F\approx0$. Therefore, we propose differential Fano factor measurement~\cite{FanoExp1, FanoExp2} could be used as a concrete toolsdf to probe the strength of twist disorder.

We finally note that MLG samples with large aspect ratio exhibit a universal maximum value of $F=1/3$~\cite{BeenakkerNoise,KatsnelsonLanduer,FanoExp1,FanoSchomerus}. For the existence of such a universal factor it is critical to have Dirac dispersions on either side of a boundary. However, as we have discussed previously, since in our case a Dirac particle on one side of the DW must become a gapped moir\'e particle on the other side (for the same $k_y$), transport across TADs do not exhibit this maximum.

\section{Conclusion}
\label{sec:Conclude}

We showed that tunneling of moir\'e electrons across TADs can be understood by the formation of an effective step potential. In general, the height of this step, $\Delta_{\mathrm{step} } = |\epsilon_\mathrm{v}^{ \mathrm{R} } - \epsilon_\mathrm{v}^{ \mathrm{L} }|$, is a function of the angle of incidence (or $k_y$) and the tilt angle of the MBZ. For normally incident particles, $\Delta_{\mathrm{step} }$ is maximized. For ballistic transport, $\Delta_{\mathrm{step} }$ results in a gap for low-energy moir\'e particles. In addition, we argue that the existence of such a low-energy gap indirectly ensures the conductance minimum of a sample is not altered by the presence of twist disorder. We propose that a peak in the differential Fano factor can be used as a diagnostic tool for twist disorder. 

In order for our results to be strictly applicable to a TBG device, one might have to account for various realistic corrections. For instance, the DWs in a TBG would always be of finite length and width, hence it might be very hard to achieve perfect reflection, since the electrons can always ``go around" the DW. Nevertheless, a careful engineering of semi-infinite twist-angle-domains can lead to interesting device applications. For instance, one can design novel ultra-low voltage [$\mathcal O(\lesssim 1 {\rm meV})$] switches by the interplay of twist-DWs and the application of a step potential by means of metallic contacts~\cite{GrapheneBridge,GHG-smoothPN}.

In a certain sense, the problem we have considered here is not very different from a tunneling problem in semiconducting 2D heterostructures. Across such junctions the band gap and the band mass changes abruptly as a result of variable doping~\cite{Morrow-EMT, SmithPhysRep96}. However, the key difference here, which drives many of the intriguing results we obtain, lies in the details of the bandstructure. For instance, the presence of a vHS in TBG provides an important scale that controls many important tunneling characteristics. Also, the ability to switch from a gapless linear dispersion to a gapped quadratic dispersion is quite unique to TBG.

In light of the above discussion it is also worth examining the limits of our 2-band model. This effective model clearly does not affix any special significance to any particular twist angle such as the magic angle. This should not be alarming since, in a way, by eliminating the higher remote bands from the full continuum model we have essentially removed the criterion to define flat bands -- the bands with very small bandwidth compared to the band gap. Therefore, strong correlation aspects aside, our results are equally applicable to the magic angle or other angles alike. In fact, later theoretical and experimental studies indeed observe interesting physics for a broad range of twist angles around the magic angle~\cite{KaxirasPRR1, KaxirasPRR2, Supermetal, MagicContWSe2}. In addition, having a finite bandwidth is a key requirement in our calculation. Had we used a minimal model that features perfectly flat bands, such as the chiral symmetric model~\cite{Tarnopolsky}, many of the tunneling phenomena we observe will cease to exist due to the lack of tunneling states. On the same note, it would be worth understanding the role of twist disorder in other multilayer graphitic systems~\cite{TLG-BS, TDBG-BS}.

Finally, we conclude by posing a few questions for future investigations. It would be interesting to understand the transport characteristic of a sample containing many (randomly distributed) domain walls~\cite{SarmaPixleyDis,AlexJason}. One can then ask whether a metal to insulator transition can be driven by increasing the number of DWs (or the strength of twist-disorder)
The $2$-band Hamiltonian we consider in this work [Eq.~\eqref{eq:2band}] bears a lot of similarities to the effective theory for the nematic transition of the interacting Bernal graphene~\cite{OskarNematic, LifshitzBLG}, merging transitions of Dirac cones in 2D crystals~\cite{MergPRA, MergPRB}, 
layered anti-ferromagnetic states in chirally stack multilayer graphene~\cite{LayerAFM}. Thus, exploring the role of DWs in such systems could also be interesting.
Given the recent developments in moir\'e materials made of transition metal dichalcogenide bilayers~\cite{WSe2/WS2, TMDFengcheng,BSmoireTMD}, it would be worthwhile to understand the effect of twist angle or strain domains on the electronic and excitonic physics of these materials as well.

\vspace{2mm}
\begin{center}
\textbf{ACKNOWLEDGEMENT}
\end{center}%
BP would like to thank Gregory Polshyn for motivation and remarks, Michael Stone for discussions in App.~\ref{sup:AdjBdy}, Chandan Setty for helpful comments. BP thanks the Pauli Center for Theoretical Studies and University of Z\"urich for funding and hospitality and acknowledges NSF DMR-1461952 for partial funding of this project. 
AT is funded from the European Union's Horizon 2020 research and innovation program under the Marie Sk\l odowska-Curie grant agreement number 701647.
TN acknowledges support from the European Research Council (ERC) under the European Union's Horizon 2020 research and innovation programme (ERC-StG-Neupert-757867-PARATOP) and from the NCCR MARVEL, funded by the Swiss National Science Foundation.
SR is supported by a Simons Investigator Grant from the Simons Foundation.


\appendix
\renewcommand{\thefigure}{\thesection \arabic{figure}}
\setcounter{figure}{0}
\renewcommand{\thesubsection}{\thesection \arabic{subsection}}
\setcounter{subsection}{0}
\renewcommand{\theequation}{\thesection \arabic{equation}}
\setcounter{equation}{0}

\section*{Appendices}

\section{The Density of States}
\label{sup:DOS}

Here we compute the density of states (DOS) corresponding to the dispersion considered in this work, i.e. Eq.~\eqref{eq:Dispersion},
\begin{align}
\rho(\epsilon) &\propto \sum_{\bs k}  \delta \left( \epsilon_k - \epsilon \right)  
= \sum_{\bs k} \epsilon_k \,  \delta  \left( \epsilon^2_k - \epsilon^2 \right)  
& (\text{for} \;\; \epsilon \geq 0) \nonumber \\
& = \int d^2 \bs k \sum_{i}^4 \frac{ \delta\left( k - k_i \right)}{ |\partial^{\phantom{2}}_k \epsilon^2_k |}
& (\text{for} \;\; \epsilon_{k_i}^2 = \epsilon^2)  \nonumber \\
&=  \int_{-\pi/2}^{\pi/2} d\phi \; \frac{\epsilon }{\sqrt{   \epsilon^2 - \epsilon_\mathrm{v}^2 \sin^2 (2 \phi - 2 \varphi )} } \nonumber \\
& =2 \pmb K \left(\epsilon_\mathrm{v}^2 / \epsilon^2 \right) .
\end{align}%
The last integral is the complete elliptic integral of the first kind, $\pmb K(z)$. We simplify this in three limits of interest. For low-energies this becomes
\begin{align}
\lim_{|\epsilon| \ll \epsilon_\mathrm{v}} \; \rho(\epsilon) \; \sim \frac{|\epsilon|}{\epsilon_\mathrm{v}} + \frac{|\epsilon|^3}{4 \epsilon^3_v} + \cdots .
\label{eq:DOSzero}
\end{align}%
The linear $\epsilon$-dependence of the leading order term is a remnant of the fact that the low-energy dispersion of the model is a Dirac dispersion. 

The logarithmic vHS in the DOS can also be observed following the expansion
\begin{align}
\lim_{\epsilon \approx \epsilon_\mathrm{v}} \; \rho(\epsilon) \sim - \log  \left \rvert \frac{\epsilon_\mathrm{v}^2}{\epsilon^2} - 1 \right \rvert +  \cdots .
\label{eq:vHSlog}
\end{align}%
Lastly, for very high energies the DOS becomes
\begin{align}
\lim_{|\epsilon| \gg \epsilon_\mathrm{v}} \;  \rho(\epsilon) \sim \pi .
\label{eq:DOSconstant}
\end{align}%

\section{Self-adjoint Matching Conditions}
\label{sup:AdjBdy}

In order to obtain the correct set of boundary conditions at the DW we may view our system analogous to a semiconductor heterojunction (such as GaAs/AlGaAs), where one typically has a band mass that abruptly changes across the junction. The Schr\"odinger equation describing coherent transport in such systems must be solved with appropriate junction conditions that takes this discontinuity in the mass into account~\cite{Ando-EMT, Morrow-EMT, Guinea-SLG-BLG, Mikito-SLG-BLG, Semenoff-Soliton}. It is important to realize that the wavefunctions in such effective band theories are not true wavefunctions (which are always continuous), rather, these are slow-varying envelope functions multiplied with (rapidly fluctuating) Bloch waves. A priori, there is no reason why such envelope functions would be continuous, exploiting that one can obtain a class of boundary conditions which respects self-adjointness of the Hamiltonian operator. We derive these conditions following~\cite{MikeBook}. In view of the eigen-equation in Eq.~\eqref{eq:EOM}, consider the following operator,
\begin{align}
\mathcal D =  \partial_x^4 - a \partial_x^2 -  i b \partial_x  + c \quad &, \quad 
(a, b, c) \in \bs R ,
\label{eq:Doperator}
\end{align}
where $a,b,c$ are defined in Eq.~\eqref{eq:defabc}. It can be explicitly verified that $\mathcal D $ is \textit{formally} self-adjoint, $\mathcal D = \mathcal D^\dag$. However, as one crosses the DW at $x=0$, the values of $(a,b,c)$ change abruptly rendering the domains of $\mathcal D$ and $\mathcal D^\dag$ unequal. We will distinguish these parameters from their left-right counterparts using $\mathrm{L, R}$ subscripts. In case of semiconducting heterojunctions, typically, have $a^{\phantom{\ast}}_{ \mathrm{L,R} } = m^\ast_{ \mathrm{L,R} }$ (band masses) and $b=0=c$.

For an arbitrary pair of normalizable wavefunction $u,v \in L^2[x_1,x_2]$, $\mathcal D$ is truly self-adjoint when
\begin{gather}
\int_{x_1}^{x_2} \left[ u^\ast \left( \mathcal D v \right)  - \left( \mathcal D u \right)^* v  \right] \, dx  \equiv Q(u,v) \Big \rvert_{x_1}^{x_2} = 0 , 
\end{gather}
In order this to be true for any value of $x_{1,2}$ on the entire space it is sufficient to ensure continuity of $Q(u,v)$ near the DW at $x=0$, $Q(u_{\mathrm{L} }, v_{\mathrm{L} }) = Q(u_{\mathrm{R} }, v_{\mathrm{R} })$, where $v_{\mathrm{L,R} } \equiv v(0^\pm)$, $v'_{\mathrm{L, R}} \equiv \partial_x v(x) \big \rvert_{x = 0^\pm } $. $Q$ can be explicitly evaluated using integration by parts,
\begin{align}
Q(u,v) = i b u^\ast v + a \left( u^\ast \partial_x v - \partial_x u^\ast v \right) + Q_0 \, ,
\label{eq:Quv}
\end{align}%
where all the boundary terms corresponding to the $\partial^4_x$ term, and $c$ are collectively denoted by $Q_0$. Since these terms remain invariant over the entire space, they match trivially at the DW. For obtaining the most general class of matching conditions for the wavefunctions we first write
\begin{gather}
\bar{\bar v}_{\mathrm{L} } = \mathcal V  \, \bar{ \bar v}_{\mathrm{R} }  \quad , \quad
\mathcal V \equiv \bem p & q \\ r & s \eem ,
\label{eq:BCmatrix}
\end{gather}%
where $\bar{\bar v}_j \equiv \left( v_j ,  v'_j \right)^T $. A similar condition for the wavefunction $u(x)$ can be written as $ \bar{\bar u}_{\mathrm{L} } = \mathcal U  \, \bar{ \bar u}_{\mathrm{R} } $. By matching the first two terms in Eq.~\eqref{eq:Quv} for left and right waves, $Q(u_{\mathrm{L} }, v_{\mathrm{L} }) = Q(u_{\mathrm{R} }, v_{\mathrm{R} })$, we first constrain $\mathcal U$. Then since such a matching must hold for an arbitrary set of wavefunctions $u, v$, we demand  $\mathcal U = \mathcal V$. All these simplify to (along with $ps - qr = a_{\mathrm{R} }/a_{\mathrm{L} }$)
\begin{gather}
q^\ast = q \quad , \quad 
s^\ast = s + \frac{i b_{\mathrm{L} }}{a_{\mathrm{L} }} q \quad , \quad p^\ast = p - \frac{i b_{\mathrm{R} }}{a_{\mathrm{R} }} q , \nonumber \\
r^\ast = - \left( \frac{i b_{\mathrm{L} }}{a_{\mathrm{L} }} p  + \frac{ b_{\mathrm{L} } b_{\mathrm{R} }}{a_{\mathrm{L} } a_{\mathrm{R} }} q + r - i \frac{i b_{\mathrm{R} }}{a_{\mathrm{R} }} s\right) .
\end{gather}%
For simplicity we fix $q=0$; evaluating $a,b$ for the left and right sides, we have,
\begin{align}
u_{\mathrm{L} } = \frac{a_{\mathrm{L} }}{a_{\mathrm{R} }} u_{\mathrm{R} } \quad , \quad
u'_{\mathrm{L} } = \frac{a_{\mathrm{L} }}{a_{\mathrm{R} }} u'_{\mathrm{R} } +  \frac i 2\left( \frac{b_{\mathrm{R} }}{a_{\mathrm{R} }} -   \frac{b_{\mathrm{L} }}{a_{\mathrm{L} }}\right) \frac{a_{\mathrm{L} }}{a_{\mathrm{R} }}  u_{\mathrm{R} } .
\label{eq:BCfinal1}
\end{align}%
For $k_y=0$ or for $\varphi_{ \mathrm{L,R} }=0$ they boil down to a simple diagonal constraint. Particularly for the later case,
\begin{align}
K_{\theta_{\mathrm{L} }}^2 u_{\mathrm{L} } = K_{\theta_{\mathrm{R} }}^2 u_{\mathrm{R} } \quad , \quad 
K_{\theta_{\mathrm{L} }}^2 u'_{\mathrm{L} } = K_{\theta_{\mathrm{R} }}^2 u'_{\mathrm{R} } .
\label{eq:BCfinal2}
\end{align}%
A discontinuous boundary condition that is purely diagonal [in the basis of $(u, u')$] may be disregarded for the purpose of computing tunneling. This is since the transfer matrix, $\mathcal M u_{\mathrm{L} } = u_{\mathrm{R} }$, would simply absorb such a factor and tunneling remains invariant under such a redefinition of $\mathcal M$.

\section{DW in Artificial Graphene: Snell's Law}
\label{sec:Snell}

This section concerns a relatively tangential scenario. Here we discuss tunneling of low-energy Dirac particles across a DW in an artificial graphene, which separates two regions with differing  Dirac speeds. As discussed in the main text, since a Dirac particle on one side always tunnels to a gapped particle on another side of the DW in a TBG system, the below discussion is not applicable to DWs in TBG. However, such a scenario may be realized in DWs in cold atom based `synthetic' graphene~\cite{GerritsmaCold1, SoltanCold2, BoadaCold3}.

The Hamiltonian describing Dirac fermions (in the pseudo-spin basis) with a relative phase or tilting $\varphi$ with respect to the DW can be obtained for $\epsilon \ll \epsilon_{\mathrm{v} }$ by linearizing $H_0$ in Eq.~\eqref{eq:2band},
\begin{align}
H_0 \simeq 2   \bem 0 & \bar q \Delta \bar K \\ q \Delta K & 0 \eem,
\end{align}%
where $q = k- \Delta K= q_x + i q_y$. The eigen-solutions of this Hamiltonian are  ($s=\pm$)
\begin{align}
\Psi_s =  \frac{1}{\sqrt 2} \begin{bmatrix}
1 \\   s i e^{- i \left(\varphi + \phi_q \right) } 
\end{bmatrix}  e^{i \bs q  \cdot \bs r} \quad , \quad \epsilon^2_q = 4 K_\theta^2 |q|^2 . 
\label{eq:LinearH0}
\end{align}%
We will view $4 K_\theta^2$ as the `Dirac speed' and we continue using the parameter $\theta$ as a proxy for tuning the speed (by means of tuning the lattice constant or the hopping parameter). Note that, had we worked with any other continuum model~\cite{LopesPRB, MCDMoire, NamKoshino17, AshvinPo, Tarnopolsky} we would have arrived at a similar low energy Hamiltonian. Except, the parameters playing the role of the Dirac speed would be different. 

\begin{figure}[t!]
\centering
\subfloat[]{ \label{fig:snell}
\includegraphics[width=0.23 \textwidth]{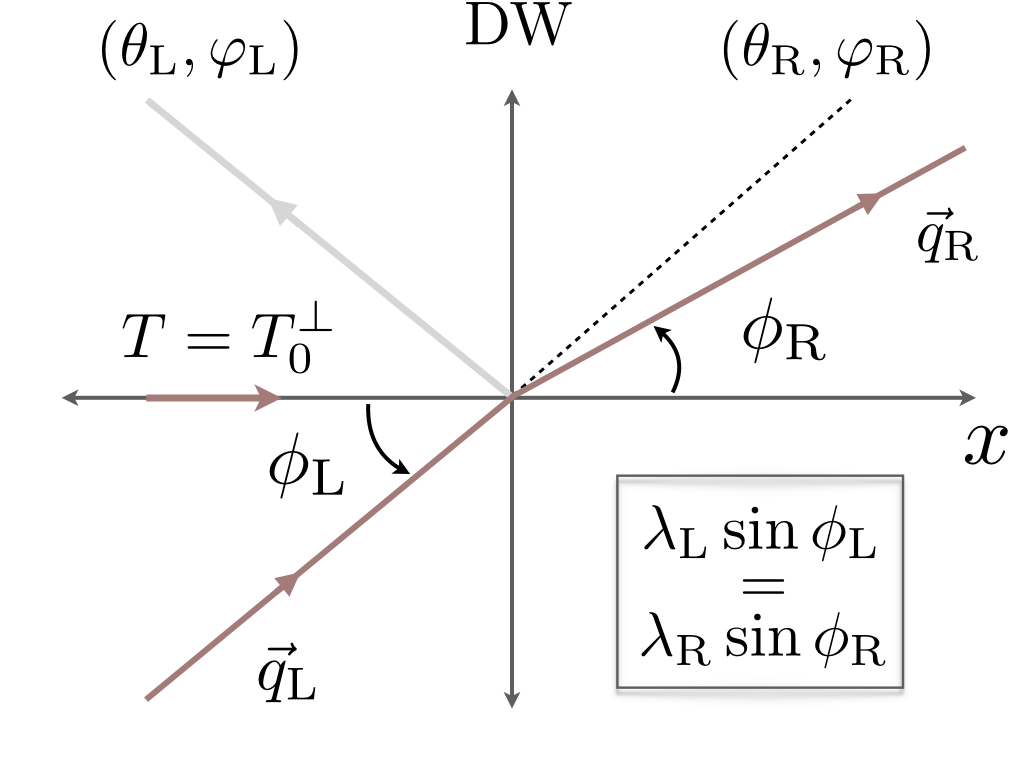} }
\subfloat[]{ \label{fig:CNPtun}
\includegraphics[width=0.245 \textwidth]{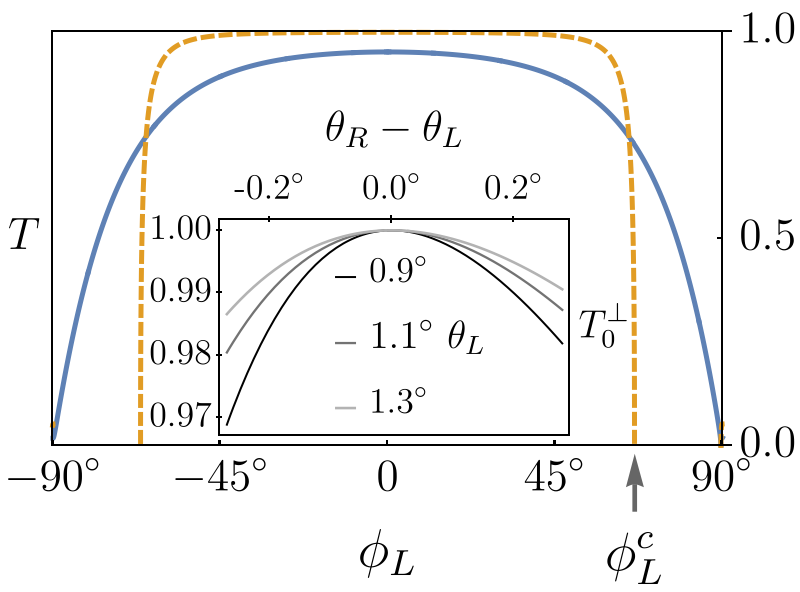}  }
\caption{
(a) Cartoon rendition of Snell's law for Dirac electrons. When $\theta_{\mathrm{R} } > \theta_{\mathrm{L} }$, for particles incident at an angle larger than a critical angle, $\phi_{\mathrm{L} } > \phi_{\mathrm{L} }^c$, there is perfect reflection (gray line). The orange (dashed) line in (b) demonstrates this for $(\theta_{\mathrm{L} }, \theta_{\mathrm{R} }) = (1.1^\circ, 1.2^\circ)$. For $\theta_{\mathrm{R} } < \theta_{\mathrm{L} }$ tunneling persists for all incidence angle, see the blue (solid) curve in (b) for $(\theta_{\mathrm{L} }, \theta_{\mathrm{R} }) = (1.1^\circ, 0.7^\circ)$. Both curves are obtained for $\epsilon = 0.1 \epsilon_\mathrm{v}^{1.1^\circ}$, however, as evident from Eq.~\eqref{eq:cnpTexp}, the dependence of $T(\phi_{\mathrm{L} })$ on energy (when finite) is extremely weak. For normal incidence ($q_y=0$), irrespective of the tilt-angle, $T=1$ for the gapless ($\epsilon=0$) particles. For low-energy gapped particles $T=T^\perp_0$ is a constant in energy. The inset in (b) shows how $T_0^\perp$ decreases from $1$ with increasing difference in twist angles.
}
\label{fig:CNPTunn}
\end{figure}

Using the energy expression in Eq.~\eqref{eq:LinearH0} one can write the following constraint for tunneling across the DW,
\begin{align}
4 K^2_{\theta_{\mathrm{L} }} \left( q_{\mathrm{L} }^2 + q_y^2 \right) = \epsilon^2 = 4 K^2_{\theta_{\mathrm{R} }} \left( q_{\mathrm{R} }^2 + q_y^2 \right) .
\label{eq:EnerCon}
\end{align} 
This allows us to write the $y$-axis momentum as $2 q_y K_{ \mathrm{L,R} } = \epsilon  \sin \phi_{ \mathrm{L,R} }$. Since momentum ($q_y$) and energy ($\epsilon$) are conserved during the tunneling process, irrespective of the value of tilt angle ($\varphi$), we obtain
\begin{align}
\lambda_{\mathrm{L} } \sin \phi_{\mathrm{L} } = \lambda_{\mathrm{R} } \sin \phi_{\mathrm{R} } ,
\end{align}
much in the spirit of the Snell's law of refraction, see Fig.~\ref{fig:snell}. Here, $\phi_{ \mathrm{L,R} }$ are incidence and transmission angles, respectively, and $\lambda_{ \mathrm{L,R} } = 2\pi/ 3K_{ \mathrm{L,R} }$ are lattice periodicities of ``graphenes" on the left and right. It is worth emphasizing that the existence of such a Snell's law solely depends on the fact that the low energy dispersion is linear. Of course, for MLG this cannot be realized since the Dirac speed in MLG fixed.

An immediate consequence of this property is, for $\theta_{\mathrm{R} } > \theta_{\mathrm{L} }$ (hence, $\phi_{\mathrm{R} } > \phi_{\mathrm{L} }$), as one increases $\phi_{\mathrm{L} }$ after some critical incidence angle, $\phi_{\mathrm{L} }^c$, $\phi_{\mathrm{R} }$ will become $\pi/2$. For any $\phi_{\mathrm{L} } > \phi^c_{\mathrm{L} }$, irrespective of $\varphi_j$, since there are no propagating modes on the right side, the electrons will simply reflect back much like total internal reflection. The associated critical angle can be obtained using $\lambda_{\mathrm{L} } \sin \phi_{\mathrm{L} }^c = \lambda_{\mathrm{R} } \sin \pi/2$ as $\phi_{\mathrm{L} }^c = \sin^{-1} \left( \lambda_{\mathrm{R} }/\lambda_{\mathrm{L} } \right)$. Of course, such a phenomenon would not exist for $\theta_{\mathrm{R} } < \theta_{\mathrm{L} }$.

In order to demonstrate the above phenomena we explicitly compute the low-energy tunneling. When there is a DW along the $y$-axis, we replace $q_x \rightarrow - i \partial_x$ and solve the wavefunctions,
\begin{align}
- \partial_x^2 \Psi +\left( q_y^2 - \frac{\epsilon^2}{4 K_\theta^2} \right) \Psi = 0. 
\label{eq:Eigen2}
\end{align}%
The solutions of this second order equation are $\pm X_j$, where $4 K^2_{\theta_j} \left( X_j^2 + q_y^2 \right) = \epsilon^2$. Therefore $X_j$ is either real or imaginary (unlike for the gapped moir\'e states which can admit both real and imaginary solutions simultaneously). The wavefunctions on the two sides of the DW are 
\bse
\begin{align}
\Psi_{\mathrm{L} } = & \bigg [ \bem \alpha_{\mathrm{L} }^+ \\ \beta_{\mathrm{L} }^+ \eem e^{i q_{\mathrm{L} } x} 
+  \bem \alpha_{\mathrm{L} }^- \\  \beta_{\mathrm{L} }^- \eem e^{- i q_{\mathrm{L} } x} \bigg ]  e^{i  y q_y }
\\
\Psi_{\mathrm{R} } = & \bem \alpha_{\mathrm{R} }^+ \\ \beta_{\mathrm{R} }^+ \eem e^{i q_{\mathrm{R} } x} e^{i  y q_y }.
\end{align}
\ese
Here, $q_j^2 = - q_y^2 + \epsilon^2/2\epsilon{_v^j}$, and $\tan \phi_j = q_y/q_j$, so $\phi_{\mathrm{L} }$ is incidence-angle and $\phi_{\mathrm{R} }$ is transmission-angle. Since incidence is only from the left we will set $|\alpha_{\mathrm{L} }^+|^2 + |\beta_{\mathrm{L} }^+|^2=1$. For the operator in Eq.~\eqref{eq:Eigen2}, comparing it with Eq.~\eqref{eq:Doperator}, we have $a_{L, R} = K_{\theta_{ \mathrm{L,R} }}^2$ and $b_{ \mathrm{L,R} }=0$. Thus, following the matching conditions in Eq.~\eqref{eq:Match} we obtain tunneling to be  
\begin{align}
T = & \; |\alpha_{\mathrm{R} }^+|^2 + |\beta_{\mathrm{R} }^+|^2 = \frac{4 q_{\mathrm{L} } q_{\mathrm{R} }}{(q_{\mathrm{L} } + q_{\mathrm{R} })^2} 
.\label{eq:TunCNP}
\end{align}
Clearly, when $\epsilon = 0$, independent of $q_y$, the tunneling reduces to $T=1$, an incarnation of Klein tunneling~\cite{NovoselovKlein, KleinEPJB11,PeresRMP}. We also note that Eq.~\eqref{eq:TunCNP} does not depend on $\varphi$. In other words, tunneling of Dirac  electrons is never affected by how the BZ is tilted with respect to the DW. 
Next, in order to understand the energy dependence of $T$, since $\epsilon \ll \epsilon_\mathrm{v}^j$, we perform the following energy expansion (for a fixed non-zero value of $q_y$)
\begin{align}
T \simeq 1 - \frac14 \frac{\Delta_{\mathrm{step} }^2 }{ (2q_y)^4}  \left( \frac{\epsilon^2}{ \epsilon_\mathrm{v}^{ \mathrm{L} } \epsilon_\mathrm{v}^{ \mathrm{R} }} \right)^2 + \mathcal O (\epsilon^6)  .
\label{eq:cnpTexp}
\end{align}%
The energy dependence of $T$ enters at $\mathcal O(\epsilon^4)$, thus $T(\epsilon) \approx 1$ for most values of energy. The correction term, which is equal to the reflectance, is proportional to $  \Delta_{\mathrm{step} }^2 = \left( \epsilon_\mathrm{v}^{ \mathrm{L} } - \epsilon_\mathrm{v}^{ \mathrm{R} } \right)^2$. This is natural since the probability of tunneling across a DW must reduce with increasing twist angle difference. For $q_y=0$, the tunneling probability reduces to a constant, $T= 4 \epsilon_\mathrm{v}^{ \mathrm{L} } \epsilon_\mathrm{v}^{ \mathrm{R} }/ \left( \epsilon_\mathrm{v}^{ \mathrm{L} } + \epsilon_\mathrm{v}^{ \mathrm{R} } \right)^2 \equiv T_0^\perp $. In the inset of Fig.~\ref{fig:CNPtun} we plot the $\theta_{ \mathrm{L,R} }$ dependence of $T_0^\perp$. 

\subsection{Born Scattering}

We now try to understand the above results, qualitatively, by re-formulating the tunneling problem in guise of an exercise in quantum mechanical  scattering~\cite{AndoAntiLoc}. In fact, we will show that, using this formalism, one can corroborate the tunneling properties discussed above with those of MLG, BLG, TBG for potential barriers.

Consider a chiral particle of dispersion, $|k|^J$, such as quasiparticles in a $J$--layered graphene. The Hamiltonian and the wavefunctions describing such particles are~\cite{McCannFalko}
\begin{gather}
\hat H_J = \begin{bmatrix} 0 & \left(\hat k^\dag \right)^J \\  \hat k^J & 0  \end{bmatrix}  \quad , \quad \nonumber
 \overlap{\bs r}{\psi_k} = e^{i \bs k \cdot \bs r} R_J  \left( \phi_k \right) \mathcal F_s(0) , \\
R_J(\phi_k) = \exp \left(- i \frac{J}{2} \phi_k \sigma_z  \right) .
\label{eq:ScattR}
\end{gather}
Note that, due to the presence of the $J/2$-spin-rotation operator, $R_J(\phi_k)$, state $\ket{\psi_k}$ gains a Berry phase of $J\pi$ after encircling a closed contour ($\phi_{k'} - \phi_{k} \equiv \delta \phi  = 2\pi$) in momentum space. When such particles scatter off of a potential, $\hat V(\bs r)$, such as by impurities or potential barriers, the angular distribution of scattering cross-section can be obtained using first order Born approximation,
\begin{align}
\Sigma_{k k'}  & \propto \left \rvert \bra{\psi_{k'} } \hat V(\bs r) \ket{\psi_k} \right \rvert^2 \sim 
|\hat V_{kk'}|^2 \cos^2  \frac{J \delta \phi_k}{2} .
\label{eq:ScatJ}
\end{align}%
Since intra-valley processes are prohibited in our model $\hat V(\bs r)$ does not act on the sublattice space, and remains diagonal. For backscattering from a barrier localized along the $k_y$--axis, that is, for $k'_x  = - k_x$ or $\phi_{k'} = \pi - \phi_k$, scattering probability becomes $\sin^2 J\phi_k $ for odd $J$, such as in MLG. Thus, for normal incidence backscattering is zero, or $T=1$. This is simply Klein tunneling~\cite{NovoselovKlein, KleinEPJB11}. For even $J$, this becomes, $\cos^2 J \phi_k$, hence, backscattering is optimal for normal incidence, as is seen in BLG~\cite{NovoselovKlein, MassiveChiral}. Although in reaching this conclusion, with the use of Eq.~\eqref{eq:ScatJ}, we implicitly assumed the strength of $\hat V(\bs r)$ to be small compared to the electron energy, one can still establish this result non-perturbatively for arbitrary strength of the barrier. This is done by performing an exact summation of the full Born series~\cite{AndoAntiLoc}. Alternative computations~\cite{GrapheneBridge,NobelSymp,RozhkovRev16} and experiments indeed confirm this result~\cite{BarrierSLGexp, FETbilayer}.
 
Before proceeding to apply the Born approximation to the tunneling in artificial graphene, we note that Eq.~\eqref{eq:ScatJ} is valid only for low energy scatterers, $\epsilon \ll |\hat V|$. As we had shown before, the strength of the potential, $\hat V(\bs r)$, is determined by $\Delta_{\mathrm{step} } \sim \mathcal{O}(\epsilon_\mathrm{v})$. Therefore, the following discussion will be restricted to the states closer to the CNP and not to those near the vHS. In order to compute $\Sigma_{kk'}$ for TBG, we recast the moir\'e wavefunction in Eq.~\eqref{eq:WaveFun} by using the rotation operator in Eq.~\eqref{eq:ScattR}. Effectively, this amounts to replacing $J \phi_k$ by the sublattice phase $ \eta_k$. For low-energy quasiparticles, $k \rightarrow q + \Delta K$, thus, $ \eta_k \simeq  \phi_q + \varphi + \pi/2$. Using this we evaluate
\begin{align}
\Sigma_{k k'}^{D} \propto |\hat V|^2 \cos^2 \frac12 \left( \eta_{k'} - \eta_k \right) \sim \Delta_{\mathrm{step} }^2 \cos^2 \frac12 \left( \phi_{q'} - \phi_q \right).
\end{align}
Firstly, $\Sigma_{k k'}^{D}$, hence tunneling, is independent of $\varphi$. Second, the amplitude of $\Sigma_{k k'}^{D}$ scales with $\Delta_{\mathrm{step} }^2$. This is in agreement with what we had evaluated in Eq.~\eqref{eq:cnpTexp}. Lastly, for backscattering, like in the case of monolayer graphene, $\Sigma_{k k'}^{D}$ vanishes for normal incidence and $T=1$. In other words, low energy transport in a TBG with a (weak) potential barrier, or for a Dirac particle in artificial graphene, is identical to that in a single layer graphene. This is consistent with the conclusions of~\cite{ChiralTunn} as well.

\section{Obtaining The Transfer Matrix}
\label{sup:DeriveM} 

In order to obtain the transfer matrix we first recast the matching conditions of Eq.~\eqref{eq:Match} in a different basis
\bse
\begin{align}
R_1 \bem p^+_{\mathrm{R} } \\ p^-_{\mathrm{R} } \eem &= E_1 \bem e_{\mathrm{L} } \\ e_{\mathrm{R} } \eem  + L_1 \bem p^+_{\mathrm{L} } \\ p^-_{\mathrm{L} } \eem 
, \\ 
R_2 \bem p^+_{\mathrm{R} } \\ p^-_{\mathrm{R} } \eem &= E_2 \bem e_{\mathrm{L} } \\ e_{\mathrm{R} } \eem + L_2 \bem p^+_{\mathrm{L} } \\ p^-_{\mathrm{L} } \eem .
\end{align}
\ese
Our goal here is to eliminate the evanescent modes, $(e_{\mathrm{L} }, e_{\mathrm{R} })^T$, and express the outgoing modes in terms of the incoming modes only. This obtains the transfer matrix as
\begin{gather}
\bem p^+_{\mathrm{R} } \\ p^-_{\mathrm{R} } \eem = \mathcal M \bem p^+_{\mathrm{L} } \\ p^-_{\mathrm{L} } \eem  , \nonumber \\
\mathcal M = \left( R_2 - E_2 E_1^{-1} R_1 \right)^{-1}  \left( L_2 - E_2 E_1^{-1} L_1 \right) .
\label{eq:TransMat}
\end{gather}%
The matrices appearing above are
\begin{align}
&E_1 = \bem 1 & - \xi \\ \chi_+ & - \xi \chi_- \eem  \,\, , \,\,
E_2 =  \bem \kappa_+ & \xi \kappa_- - \zeta \\ \kappa_+ \chi_+ & (\xi \kappa_- - \zeta) \chi_- \eem 
\nonumber , \\
&L_1 = \bem 1 & 1 \\ e^{i \eta_{\mathrm{L} }^+} & e^{i \eta_{\mathrm{L} }^-} \eem  \,\, , \,\,
L_2 = i k_{\mathrm{L} } \bem 1 & - 1 \\  e^{i \eta_{\mathrm{L} }^+} & - e^{i \eta_{\mathrm{L} }^-} \eem 
, \\
&R_1 = \xi \bem 1 & 1 \\ e^{i \eta_{\mathrm{R} }^+} & e^{i \eta_{\mathrm{R} }^-} \eem  \,\, , \,\,
R_2 = i \xi k_{\mathrm{R} } \bem  1 &  - 1 \\   e^{i \eta_{\mathrm{R} }^+} & - e^{i \eta_{\mathrm{R} }^-} \eem + \frac{\zeta}{\xi} R_1 
\nonumber .
\end{align}

Independent of the transfer matrix formalism, one could also obtain tunneling coefficients by making use of the (probability) current operator (for $\varphi=0$),
\begin{align}
J_x = - i \Psi^\dag \bem 0 & \partial_x + 2 k_y \\ \partial_x - 2 k_y & 0 \eem \Psi + {\rm h.c.} .
\end{align}%

\section{Tunneling and Quantum Noise}
\label{sup:Noise}

Electric current inside a conductor is proportional to the density of charge carriers, $n_e$. If temperature is  high enough this number fluctuates following Boltzmann distribution, $\delta n = n_e - \expect{n_e} \neq 0$. This manifests in the linear response function or conductance, due to fluctuation--dissipation theorem, where fluctuation in the current is $\delta I = I - \expect{I}$. This is regarded as the thermal noise, which, in some sense, does not carry more information than the steady state conductance itself, $\expect{I}$, since it simply is a measure of temperature.

For our discussion we are interested in the so-called \textit{shot noise}, that originates from quantum mechanical fluctuations in charge carriers. This can be used, unlike thermal noise, to probe the transport or non-equilibrium states of a conductor, even at zero temperature. Consider, for instance, the tunneling problem across a barrier or a DW. We can characterize the incident state by an occupation number $n_i$ (zero or one). Similarly, occupation for the reflection state and tunneling state are $n_{r}$ and $n_t$, respectively. If we repeat the tunneling experiment several times, $\expect{n_i} = 1, \expect{n_r} = R, \expect{n_t} = T$. Not for finite temperature, all of them get multiplied with the Fermi function, $f(\epsilon)$. Since in every instance of this experiment a particle is either reflected or transmitted, we have, $\expect{n_t n_{r}}=0$. Using similar arguments, one can obtain the correlation between the transmitted and the reflected beams as $\expect{\delta n_t \delta n_r} = - T R$, and $\expect{\delta n_t^2} = \expect{\delta n_r }^2 = T R$. These 2-point functions are called partition noises as the barrier essentially partitions the incident beam into either a reflected or a tunneled beam. Clearly, partition noise is maximum for $T=1/2$ and vanishes for $T=1$ or $R=1$.

A natural formulation of the above quantities (noises) pertaining to transport measurements can be done through currents in different channels,
\begin{align}
I_c &= \frac{e}{h} \int n_c (\epsilon) d\epsilon  \quad (c=i,r,t) , 
\\
\expect{I_c} &= \frac{e}{h} \int f(\epsilon) N_c d\epsilon \quad (N_{i,r,t} = 1, R, T).
\label{eq:NoiseCurr}
\end{align}%
The aforementioned noises can thus be measured (strictly speaking for low frequency fluctuations) by the correlation function $\expect{\delta I_{c1} \delta I_{c2} } $. A quantity of experimental interest is the current `noise-power', defined for a pair of channels as
\bse
\begin{align}
S_{c1 c2} = {\rm g}  e & \expect{\delta I_{c1}  \delta I_{c2} }  = \frac{{\rm g}  e}{h} \int \expect{\delta n_{c1} \delta n_{c2} }  d \epsilon , \\ 
S_{tt} &= \frac{{\rm g}  e}{h} \int T f ( 1- T f) d \epsilon  , \\
S_{rr} &= \frac{{\rm g}  e}{h} \int R f \left( 1- R f \right) d \epsilon  , \\
S_{rt} &=  - \frac{{\rm g}  e}{h} \int T f R f d \epsilon   .
\label{eq:NoisePowers}
\end{align}%
\ese
Here ${\rm g}$ is a symmetry factor; for spin-1/2 particles $\rm g = 2$. We will use ${\rm g} = 4$ since we  also have valley symmetry. The above expression for $S_{tt}$ can also be derived using a Poissonian distribution of time intervals between the arrival times of the particles at the barrier. When $S_{tt}$ approaches its maximum value,  ${\rm g} e \expect{I} \equiv S^0_{tt} $, called Poisson noise or Schottky noise, it signals an uncorrelated arrival at the barrier. In order to measure whether the transport is maximally noisy (Poissonian) or not (sub-Poissonian), naturally, one can make use of their ratio, 
\begin{align}
F = \frac{S_{tt}}{S_{tt}^0}  = \frac{\int T f ( 1- T f) d \epsilon }{\int T f d \epsilon } \leq 1 .
\end{align}%
This is the Fano factor, see also Eq.~\eqref{eq:FanoDef}. Note that it is the $(1-Tf)$ factor that essentially drives the system from noisy to noise-free transport. For instance, in ballistic systems ($T=1$) the tunneling shot noise will vanish as temperature approaches zero, thus $F \approx 0$. With increasing temperature, the thermal noise may dominate over the shot noise; however, for very high temperature, since $1 - f \approx 1$, Poisson noise is recovered, $F \approx 1$. On the other hand for a diffusive system accompanied by very small transparency ($T \ll 1$) the tunneling noise could be Poissonian even for zero temperature, $F \approx 1$. Thus, $F$ provides key insight into the possible mechanism of transport in a conductor.

In summary, at zero temperature $F=1$ means transport is noisy and diffusive (such as in a disorder-free metal). If $F < 1$ there are open quantum channels which can allow ballistic transport (such as a disordered metal~\cite{DiffusiveMetal}). Of course, $F=0$ is a noiseless ballistic transport, such as the classical dynamics of Dirac fermions.

\end{document}